\documentclass[12pt,oneside,english]{amsart}
\usepackage{ae,aecompl}

\usepackage[T1]{fontenc}
\usepackage[latin9]{inputenc}
\usepackage[verbose,tmargin=1.24in,bmargin=1.24in,lmargin=1.24in,rmargin=1.24in]{geometry}

\makeatletter
\renewenvironment{abstract}{%
  \ifx\maketitle\relax
    \ClassWarning{\@classname}{Abstract should precede title in AMS style}%
  \fi
  \global\setbox\abstractbox=\vtop \bgroup
    \normalfont\Small
    \list{}{\labelwidth\z@
            \leftmargin0.5in 
            \rightmargin0.5in 
            \listparindent\normalparindent
            \itemindent\z@
            \parsep\z@ \@plus\p@
            
    }%
    \item[\hskip\labelsep\scshape\abstractname.]%
}{%
  \endlist\egroup
  \ifx\@setabstract\relax \@setabstracta \fi
}
\makeatother

\usepackage{csquotes}
\usepackage{color}
\usepackage{babel}
\usepackage{amstext}
\usepackage{amsthm}
\usepackage{amssymb}
\usepackage{setspace}
\usepackage[authoryear]{natbib}
\usepackage[unicode=true,pdfusetitle,
 bookmarks=true,bookmarksnumbered=false,bookmarksopen=false,
 breaklinks=false,pdfborder={0 0 0},pdfborderstyle={},backref=false,colorlinks=true]
 {hyperref}
\hypersetup{citecolor=blue, colorlinks=true, linkcolor=blue}

\makeatletter
\let\old@footnotetext\@footnotetext
\renewcommand\@footnotetext[1]{%
  \begingroup
    \interlinepenalty\@M
    \old@footnotetext{#1}%
  \endgroup
}
\makeatother

\makeatletter
\numberwithin{equation}{section}
\numberwithin{figure}{section}
\theoremstyle{remark}
\newtheorem{rem}{\protect\remarkname}
\theoremstyle{plain}
\newtheorem{thm}{\protect\theoremname}
\newtheorem{cor}{\protect\corollaryname}
\theoremstyle{definition}
\newtheorem{example}{\protect\examplename}
\ifx\proof\undefined\
  \newenvironment{proof}[1][\proofname]{\par
    \normalfont\topsep6\p@\@plus6\p@\relax
    \trivlist
    \itemindent\parindent
    \item[\hskip\labelsep
          \scshape
      #1]\ignorespaces
  }{%
    \endtrivlist\@endpefalse
  }
  \providecommand{\proofname}{Proof}
\fi
\theoremstyle{plain}
\newtheorem{lem}{\protect\lemmaname}

\usepackage{babel}

\usepackage{refcount}
\usepackage{etoolbox}

\newcounter{assumptioncount}
\newcounter{displayassumptioncount}

\newtheorem{assumptionenv}{Assumption}

\renewenvironment{assumptionenv}[1][]{%
  \stepcounter{assumptioncount}%
  \ifnum\value{assumptioncount}=1
    \def\@currentlabel{1(p)}%
    \label{ass:Assumption1p}%
  \else
    \ifnum\value{assumptioncount}=2
      \def\@currentlabel{1(q)}%
      \label{ass:Assumption1q}%
    \else
      \setcounter{displayassumptioncount}{\numexpr\value{assumptioncount}-1\relax}%
      \def\@currentlabel{\arabic{displayassumptioncount}}%
      \ifnum\value{displayassumptioncount}>1
        \label{ass:Assumption\arabic{displayassumptioncount}}%
      \fi
    \fi
  \fi
  \ifstrempty{#1}{%
  }{%
    \label{#1}%
  }%
  \begin{trivlist}
  \item[\hskip\labelsep
        \textbf{Assumption \@currentlabel}]%
}{%
  \end{trivlist}
}

\let\oldref\ref
\renewcommand{\ref}[1]{%
  \ifstrequal{#1}{ass:Assumption1p}{1(p)}{%
    \ifstrequal{#1}{ass:Assumption1q}{1(q)}{%
      \ifstrequal{\oldref{#1}}{1}{Assumption 2}{%
        \ifstrequal{\oldref{#1}}{2}{Assumption 3}{%
          \ifstrequal{\oldref{#1}}{3}{Assumption 4}{%
            Assumption~\oldref{#1}%
          }%
        }%
      }%
    }%
  }%
}

\newcommand{\refpq}[2]{%
  \ifstrequal{#1}{ass:Assumption1}{%
    \ifstrequal{#2}{p}{\hyperlink{ass:Assumption1p}{\ref{#1}(p)}}{%
      \ifstrequal{#2}{q}{\hyperlink{ass:Assumption1q}{\ref{#1}(q)}}{%
        \ref{#1}(\texttt{#2})%
      }%
    }%
  }{%
    \ifstrequal{#1}{ass:cond_nonsingularity}{%
      \ifstrequal{#2}{p}{\hyperlink{ass:cond_nonsingularity_p}{\ref{#1}(p)}}{%
        \ifstrequal{#2}{q}{\hyperlink{ass:cond_nonsingularity_q}{\ref{#1}(q)}}{%
          \ref{#1}(\texttt{#2})%
        }%
      }%
    }{%
      \ref{#1}(\texttt{#2})%
    }%
  }%
}

\newcommand{\refp}[1]{\refpq{#1}{p}}
\newcommand{\refq}[1]{\refpq{#1}{q}}

\def\ind{\!\perp\!\!\!\perp}

\providecommand{\examplename}{Example}
\providecommand{\remarkname}{Remark}
\providecommand{\corollaryname}{Corollary}
\providecommand{\theoremname}{Theorem}
\providecommand{\lemmaname}{Lemma}

\date{\today}
\thanks{\texorpdfstring{$^\dag$ Department of Economics, MIT}
 }
\thanks{\texorpdfstring{$\S$ Department of Economics, University of Bristol and University of Melbourne}
}
\thanks{\texorpdfstring{
We are grateful to Jean-Pierre Florens, Michal Koles\'ar, Martin Weidner, Frank Windmeijer for useful comments. We also thank  seminar participants at the 2019 
Bristol-TSE Econometrics Workshop, the 2024 Microeconometrics and Policy Evaluation Barcelona Workshop, the 2024 Groningen Workshop on Causal Inference and Machine Learning, CUHK, University of Melbourne and University of Oxford for useful comments. 
This paper replaces the manuscript previously circulated under the title `Heterogenous Coefficients, Discrete Instruments, and Identification of Treatment Effects'}
}
\title{\vspace{0em}  
Identification of Treatment Effects under Limited Exogenous Variation}

\author[Whitney K. Newey and Sami Stouli]{Whitney K. Newey$^{\dag}$ and Sami Stouli$^{\mathsection}$}

\begin{document}

\vspace*{0em}  
\maketitle
\thispagestyle{empty}  

\vspace{0em}  

\begin{abstract}
Multidimensional heterogeneity and endogeneity are important features of a wide class of econometric models. 
With control variables to correct for endogeneity, nonparametric identification of treatment effects requires strong support conditions. 
To alleviate this requirement, we consider varying coefficients specifications for the conditional expectation function of the outcome given a treatment and control variables. 
This function is expressed as a linear combination of either known functions of the treatment, with unknown coefficients varying with the controls, or known functions of the controls, with unknown coefficients varying with the treatment. 
We use this modeling approach to give necessary and sufficient conditions for identification of average treatment effects. 
A sufficient condition for identification is conditional nonsingularity, that the second moment matrix of the known functions given the variable in the varying coefficients is nonsingular with probability one. 
For known treatment functions with sufficient variation, we find that triangular models with discrete instrument cannot identify average treatment effects when the number of support points for the instrument is less than the number of coefficients. 
For known functions of the controls, we find that average treatment effects can be identified in general nonseparable triangular models with binary or discrete instruments. 
We extend our analysis to flexible models of increasing dimension and relate conditional nonsingularity to the full support condition of \citet{Imbens Newey 2009}, thereby embedding semi- and non-parametric identification into a common framework. 
\end{abstract}


\textsc{\small{}Keywords:}{\small{} Identification; Conditional nonsingularity; Limited exogenous variation; Treatment effects; Heterogeneous coefficients; Control variable; Discrete instruments; Testability.}{\small\par}

\thispagestyle{empty}

\section{Introduction}

Nonseparable and/or multidimensional heterogeneity is important. It is present in discrete choice models as in \citet{McF1973} and \citet{HW1978}.
Multidimensional heterogeneity in demand functions allows price and income elasticities to vary over individuals in unrestricted ways, e.g., \citet{HN2016} and \citet{KS2018}. It allows general variation in production technologies. Treatment effects that vary across individuals require intercept and slope heterogeneity.

Endogeneity is often a problem in these models because we are interested in the effect of an observed choice, or treatment variable on an outcome and the choice or treatment variable is correlated with heterogeneity.
Control variables provide an important means of controlling for endogeneity with multidimensional heterogeneity. A control variable is an observed
or estimable variable that makes heterogeneity and treatment independent when it is conditioned on. The conditional cumulative distribution function (CDF) of a choice variable given an instrument can serve as a control variable in triangular models (\citet{Imbens Newey 2009}).

In fully nonparametric and nonseparable models, identification of average or quantile treatment effects requires a full support condition, that the support of the control variable conditional on the treatment variable is equal to the marginal support of the control variable.
This restriction is often not satisfied in practice; e.g., see \citet{Imbens Newey 2009} for Engel curves. In triangular models the full support condition cannot hold
when all instruments are discrete and the treatment variable is continuous.

One approach to this problem is to focus on identified sets for objects of interest, as for quantile effects in \citet{Imbens Newey 2009}.
Another approach is to consider restrictions on the model that allow for point identification. In a triangular model with continuous treatment, \citet{Flo Heck Meg Vytl 2008} gave identification results when the outcome equation is a polynomial in the endogenous variable, and \citet{MT:2016}
when the outcome equation is a linear combination of known transformations of the endogenous variable that are not necessarily polynomials. \citet{Torgo 2015} and \citet{Fev Hault 2015} gave identification results when there is only scalar heterogeneity in the outcome equation.

In this paper we give identification results when the Control Regression Function (CRF), the regression function of the outcome given the treatment and the controls, is a linear combination of either known functions of the treatment, with unknown coefficients varying with the controls, or known functions of the controls, with unknown coefficients varying with the treatment.  
We further assume that the Average Structural Function (ASF, \citet{Blundell Powell 2003}), the outcome structural function when heterogeneity has been integrated out, coincides with the CRF integrated over the controls. A generic sufficient condition for identification of average treatment effects in our models is that the 
second moment matrix of the
known functions given the variable in the varying coefficients 
is nonsingular with probability one. This framework is a generalization of heterogeneous coefficients formulations where the outcome function is linear in known functions of the  treatment, and where the coefficients in this linear combination are mean independant of heterogeneity given the controls.

A main benefit of this modeling framework is that it allows for identification of average treatment effects under limited exogenous variation, i.e., without full support.
For important cases of practical relevance, this provides a strong
motivation for placing restrictions on the way either the treatment or the control variables affect the outcome, so that point identification is preserved. Leading cases include continuous treatment with conditional support varying across control variable values, continuous treatment in a triangular model with discrete
instruments, and multiple treatments without common support or strong overlap. 
For all these cases, our flexible formulations can deliver point identification. We obtain these results from the assumed varying coefficients structure of the CRF.

We specialize our results to average treatment effects in triangular models with discrete instruments, and we extend the analysis to quantile and distributional treatment effects. 
For known treatment functions with sufficient variation, we find that a necessary condition for ASF identification is that the number of support points of the discrete instruments is at least as large as the number of coefficients. 
With known functions of control variables, we find that identification can be achieved with binary or discrete instruments in general nonseparable triangular models when restrictions are placed on how control variables can affect heterogeneity.

These results extend \citet{Flo Heck Meg Vytl 2008} in allowing for outcome structural functions that are linear in nonpolynomial functions of the treatment variable, and in allowing for discrete instruments. 
We also take a different approach to identification, focusing here on conditional nonsingularity of second moment matrices instead of measurable separability. 
These results also generalize those of \citet{MT:2016} to allow for both estimable and observable control variables in the modeling of treatment effects. 
Our results generalize the heterogeneous coefficients formulations in both papers, to allow for a class of outcome structural functions that are not necessarily linear in known functions of the treatment, 
and for CRFs that are linear in known functions of the controls given the treatment.  
We also go beyond average effects by extending the identification analysis to quantile and distributional treatment effects. 
In addition, our modeling framework generalizes that of \citet{NeweyStouli2021} to allow for the CRF to be a linear combination of known functions of only one of either the treatment or the controls, rather than known functions of each. While this literature focuses on sufficient conditions for identification, 
we study necessity as well as sufficiency, we extend the identification analysis to flexible models of increasing dimension, and we establish testability of our model restrictions.

\begin{sloppy}
A first main contribution of our paper is the formulation of a unified
framework for identification of treatment effects under conditional nonsingularity for a large class
of models, some of which had been considered separately in the previous
literature. Our analysis reveals that models as distinct as linear
quantile regression control variable models (e.g., \citet{MaKo:2006},
\citet{Lee:2007}, \citet{Jun:2009}, \citet{CFNSV:2020}), binary
and multiple treatment effects models (\citet{RR:1983}, \citet{Imb:2000},
\citet{NeweyStouli:2022}), and heterogeneous coefficients models
with continuous treatment (\citet{Flo Heck Meg Vytl 2008}, \citet{MT:2016}),
are all members of a larger class that is amenable to a unified treatment of identification,  which then applies to each individual model. 
\par\end{sloppy}

Second, we give identification conditions that are necessary as well as sufficient for the ASF.
Necessity is important in order to demonstrate testability of identification (e.g., \citet{Breusch:1986}). 
Conditions that are both necessary and sufficient are important for the determination of minimal conditions for identification. 
We are thus able to characterize minimal conditions for a very large class of models. 
In triangular models with discrete instruments, these results allow us to establish that the number of coefficients cannot be larger than the number of support points for the instrument. 
These results generalize those in \citet{NeweyStouli:2022} from discrete to general treatments, from observable to general control variables, and from CRFs that are linear in known functions of the treatment to CRFs that are linear in known functions of either the treatment or the controls, but not both.

Third, we show that average treatment effects are identified in general nonseparable models under limited exogenous variation, when restrictions are placed on how control variables can affect the 
CRF. 
For general nonseparable triangular models with binary or discrete instruments, this means that identification can be achieved when restrictions are placed on the relationship between heterogeneity and control variables in the model.  
This formulation alleviates the full support requirement and provides a novel approach for the modeling of treatment effects in the presence of multidimensional heterogeneity and discrete instruments.

Fourth, we extend the identification analysis to models of increasing dimension, by establishing a connection between semi- and non-parametric identification conditions for treatment effects. 
When conditional nonsingularity holds for each element of an increasing sequence of suitable approximating functions, we show that the full support condition is satisfied, and hence average treatment effects are nonparametrically identified. 
Thus our modeling approach provides an encompassing framework for identification of treatment effects, from identification under limited exogenous variation in models of fixed dimension to nonparametric identification under full support.

\begin{sloppy}
Fifth, we establish testability of our model specifications (e.g., \citet{KR1950}), and characterize the complete set of implied testable restrictions.  
For triangular models, these results hold in the presence of multidimensional heterogeneity and a discrete valued instrument, in this way going beyond \citet{Chesher 2007}, \citet{Stouli:2012},
\citet{Fev Hault 2015}, \citet{Torgo 2015}, and \citet{MT:2016}.
Our results do not require full conditional independence of heterogeneity and instruments 
given the control variables, complementing the analysis in \citet{DHS:2024} 
on testability of the exclusion restriction in nonseparable triangular 
models with stronger independence conditions. 
Compared to this previous literature on continuous treatment effects, our analysis of necessary conditions for identification and our characterization of implied testable restrictions are, to the best of our knowledge, novel.
\par\end{sloppy}

In Section \ref{sec:Section2} we introduce the modeling framework. 
Section \ref{sec:Section3} gives the identification analysis and Section \ref{sec:Section4} discusses conditional nonsingularity. 
In Section \ref{sec:Section5} we specialize our results to triangular models.
Section \ref{sec:Section6} gives results on model testability. Section \ref{sec:Section7} discusses estimation, and Section \ref{sec:Section8} concludes. 
All proofs are given in the Appendix.

\section{Modeling framework} \label{sec:Section2}

Let $X$ denote an endogenous treatment, and $\varepsilon$ a structural disturbance vector of unrestricted dimension, with CDF $F_{\varepsilon}$. 
A general nonseparable treatment effects model for an outcome variable $Y$ is
\begin{equation} \label{eq:g(x,e)}
Y=g(X,\varepsilon),\quad\varepsilon\ind X\mid V,
\end{equation}
where $\varepsilon$ is independent of $X$ conditional on $V$, an observable or estimable control variable with CDF $F_{V}$. 
For this model, 
a leading structural object of interest is the ASF,
\[ 
    \mu(X) \equiv \int g(X,\varepsilon) dF_{\varepsilon}(\varepsilon).
\]

In addition to conditional independence in (\ref{eq:g(x,e)}), 
an assumption required for ASF identification is full support, 
that the conditional support of $X$ given $V$ is the same as the
marginal support of $X$ (\citet{Imbens Newey 2009}). 
The ASF can then be expressed in terms of observable or estimable quantities (\citet{Blundell Powell 2003}):
\begin{equation} \label{eq:integrated_CRF}
    \mu(X) = \int E[Y \mid X,V=v] dF_{V}(v). 
\end{equation}
Full support allows for this integral to be well-defined, and hence for ASF 
identification as a known functional of the CRF $E[Y | X, V]$. 
Full support requires the treatment to have unrestricted variation once the control variable is conditioned on.

A common occurrence in practice is that the full support assumption
does not hold. This means that, with positive probability, the support of $X$ given $V$ is only a strict subset of the marginal support of $X$, i.e., the treatment has limited exogenous variation. 
To accommodate this feature, the CRF in (\ref{eq:integrated_CRF}) constitutes a natural modeling target, on which restrictions can be placed to alleviate the full support requirement. 

In this paper, we introduce a modeling framework that allows for two types of restrictions on the CRF, each corresponding to a distinct class of flexible models, and each allowing for identification of treatment effects under limited exogenous variation. Specifically, we place restrictions on how either $X$ or $V$ affects the CRF, which we specify as either a linear combination of known functions of $X$, or a linear combination of known functions of $V$, but not both.

\subsection{CRFs with known functions of $X$}

In the first class of models we consider, the CRF is a linear combination of known functions $p(X)$, of finite and known dimension $J$, with 
varying coefficients that are unknown functions of $V$.
\begin{assumptionenv}\label{ass:Assumption1p}
Given a random vector $(Y,X,V)'$, 
the CRF takes the form 
\begin{equation} \label{eq:controlregression_1}
    E[Y|X,V]=p(X)'q_{0}(V),
\end{equation}
for known functions $p(X)$ and unknown functional coefficients $q_{0}(V)$. 
\end{assumptionenv}

Under Assumption \ref{ass:Assumption1p}, while the functional relationship between the treatment and the CRF is effectively restricted to belong to a known class of functions, the way the treatment affects the CRF is
unrestricted within that class across control variable values. 
For example, in the simplest case $p(X)=(1,X)'$, the implied CRF specification $E[Y|X,V]=q_{01}(V)+q_{02}(V)X$ imposes linearity of the CRF in $X$, while allowing the intercept
and slope coefficients to vary freely across values of $V$.

A leading example of a model that satisfies Assumption \ref{ass:Assumption1p} is 
a heterogeneous
coefficients formulation of $g(X,\varepsilon)$ in (\ref{eq:g(x,e)}), of the form
\begin{equation} \label{eq:hcoefs_model_p}
Y=p(X)^{\prime}\beta(\varepsilon),
\end{equation}
where the coefficients vector $\beta(\varepsilon)$ 
is mean independent of the endogenous variable $X$ conditional on $V$:  
\begin{equation} \label{eq:CMI}
E[\beta(\varepsilon)\mid X,V]=E[\beta(\varepsilon)\mid V]. 
\end{equation}
This class of models (\ref{eq:hcoefs_model_p})-(\ref{eq:CMI}) is one where 
$X$ is known to affect the CRF only through a vector of known functions $p(X)$. 
Conditional mean independence property (\ref{eq:CMI}) and the form of the outcome function $p(X)'\beta(\varepsilon)$ in (\ref{eq:hcoefs_model_p}) together
imply that: 
\begin{equation*} 
E[Y|X,V]=p(X)^{\prime}E[\beta(\varepsilon)|X,V]=p(X)^{\prime}E[\beta(\varepsilon)|V]=p(X)^{\prime} q_{0}(V),  q_{0}(V)\equiv E[\beta(\varepsilon)|V], 
\end{equation*}
with unknown coefficients $q_{0}(V)$. 
This specification places no restrictions on 
$E[\beta(\varepsilon)|V]$, 
and hence does not restrict how control variables can affect the 
coefficients. This is a generalization of  
\citet{Flo Heck Meg Vytl 2008} to
allow $p(X)$ to be any functions of $X$ rather than just powers of $X$. This restricted nonparametric regression is of the varying coefficients type considered by \citet{Cai:2006}. 

An example for $p(X)$ known 
naturally arises 
with discrete treatment. 
A general formulation is to let $X$ be a vector of dummy variables $X(t)$, $t\in\{1,\ldots,T\}$,
taking value one if treatment regime $t$ occurs and zero otherwise,
and setting
\begin{equation}
p(X)=(1,X(1),\ldots,X(T))'.\label{eq:p(X)}
\end{equation}
This formulation generalizes the \citet{RR:1983} binary treatment effects
model to multivalued and/or multiple treatments, and is not restrictive
when using mutually exclusive treatment regimes in the definition
of $X$. This formulation also relaxes the conditional independence assumption commonly imposed for identification of treatment effects, an assumption that is not necessary for identification and can be replaced by the weaker conditional mean independence property (\ref{eq:CMI}).\footnote{See
Section 4 in \citet{NeweyStouli:2022} for a detailed 
analysis and an equivalent formulation of the conditional mean independence assumption in terms of potential outcomes.}

\subsection{CRFs with known functions of $V$} \label{subsec:CRF_knownq}

In the second class of models we consider, the CRF is a linear combination of known functions $q(V)$, of finite and known dimension $K$, with varying coefficients that are unknown functions of $X$.
\begin{assumptionenv} \label{ass:Assumption1q}
Given a random vector $(Y,X,V)'$, the CRF takes the form 
\begin{equation} \label{eq:controlregression_2}
E[Y|X,V]=p_{0}(X)'q(V),  
\end{equation}
for known functions $q(V)$ and unknown functional coefficients $p_{0}(X)$.
\end{assumptionenv}
Under Assumption \ref{ass:Assumption1q}, while the functional relationship between the control variables and the CRF is effectively restricted to belong to a known class of functions, the way control variables affect the CRF is unrestricted within that class across treatment values. For example, in the simplest case $q(V)=(1,V)'$, the implied CRF specification $E[Y|X,V]=p_{01}(X)+p_{02}(X)V$ imposes linearity of the CRF in $V$, while allowing the intercept and slope coefficients to vary freely across values of $X$.

With $X$ continuous, a leading example of a model that satisfies Assumption \ref{ass:Assumption1q} is a heterogeneous
coefficients representation of (\ref{eq:g(x,e)}), of the form
\begin{equation} \label{eq:hcoefs_model_p*}
Y=p^{*}(X)^{\prime}\beta(\varepsilon), 
\end{equation}
where $\varepsilon \ind X | V$ and $p^{*}(X)$ is a vector of unknown functions of arbitrarily large dimension, and where 
$E[\beta(\varepsilon)|V]$ is a vector of linear combinations of known functions
\begin{equation} \label{eq:cond mean het coefs}
E[\beta(\varepsilon)\mid V]=\Omega'q(V),
\end{equation}
for an unknown $K\times J$ matrix $\Omega$. 
When $p^{*}(X)$ is a vector of approximating functions such
as splines or wavelets, this model can be viewed as an approximation
to the general nonseparable model (\ref{eq:g(x,e)}) where $\beta(\varepsilon)$
are varying coefficients in an expansion of $g(X,\varepsilon)$ in
$p^{*}(X)$, as in \citet{HN2016}.
With $p^{*}(X)$ a vector of unknown functions of arbitrarily large dimension, the outcome function $g(X,\varepsilon)$ is effectively unrestricted under (\ref{eq:hcoefs_model_p*}).\footnote{For $x\mapsto g(x,\varepsilon)$ a smooth function uniformly in $\varepsilon$ and with $p^{*}(X)$ chosen from a suitable class of approximating functions, there is $\beta(\varepsilon)$ that depends on $J$ such that $E[\{g(X,\varepsilon)-p^{*}(X)'\beta(\varepsilon)\}^{2}]\rightarrow0$
as $J\rightarrow\infty$. 
This is because the (uniform) approximation error for $x\mapsto g(x,\varepsilon)$ is bounded for each value of $\varepsilon$ (e.g., \citet{Powell:1981}), and hence uniformly over $\varepsilon$ if $x\mapsto g(x,\varepsilon)$ is smooth uniformly
in $\varepsilon$. 
The case where $Y$ is binary or discrete can be accommodated if there is $\rho$ such that $Y=g(X,\varepsilon)+\rho$ and $E[\rho|X,V]=0$ and $x\mapsto g(x,\varepsilon)$ is smooth uniformly in $\varepsilon$. This formulation extends representations proposed in \citet{CDGHN:2025} to the control variable case. \label{fn:approximation}} 
In this example, restrictions are thus placed on the way control variables affect  $\beta(\varepsilon)$, rather than on $g(X,\varepsilon)$.

This class of models (\ref{eq:hcoefs_model_p*})-(\ref{eq:cond mean het coefs}) 
is one where $V$ is known to affect the CRF only through a vector of
known functions $q(V)$. 
Conditional independence and the form of the structural function $p^{*}(X)'\beta(\varepsilon)$ in (\ref{eq:hcoefs_model_p*}) together
imply that:
\begin{align*} 
E[Y|X,V] & =p^{*}(X)^{\prime}E[\beta(\varepsilon)|X,V]=p^{*}(X)^{\prime}E[\beta(\varepsilon)|V]=p^{*}(X)^{\prime}\left\{ \Omega'q(V)\right\} \nonumber \\
 & = \{\Omega p^{*}(X)\}'q(V)= p_{0}(X)^{\prime}q(V),\quad p_{0}(X)\equiv\Omega p^{*}(X),  
\end{align*}
with unknown coefficients $p_{0}(X)$. 
Compared to (\ref{eq:hcoefs_model_p})-(\ref{eq:CMI}), 
specification (\ref{eq:hcoefs_model_p*})-(\ref{eq:cond mean het coefs}) generalizes the outcome function to allow for general nonseparable models, but restricts the way control variables can affect the CRF, thereby defining an alternative flexible varying coefficients structure for the CRF. 

An example for $q(V)$ known naturally arises with discrete control variables $V$. 
A general formulation is to let $V$ be a vector of dummy variables $V(s)$, $s\in\{1,\ldots,S\}$, taking value one if control value $s$ occurs and zero otherwise,
and setting
\[
\Omega\in\mathbb{R}^{(S+1)\times J},\quad q(V)=(1,V(1),\ldots,V(S))'.
\]
This formulation allows for unrestricted modeling of $E[\beta(\varepsilon)|V]$ when using mutually exclusive control regimes in the definition of $V$.

\begin{rem}
Additional exogenous covariates $Z_{1}$ can be incorporated straightforwardly in CRF models (\ref{eq:controlregression_1}) and (\ref{eq:controlregression_2}), either through the known CRF component or through the unknown component. 
We illustrate briefly for the case of the known component, a convenient choice in practice. With covariates $Z_{1}$, the CRF (\ref{eq:controlregression_1}) takes the form 
\[
E\left[Y\mid X,Z_{1},V\right]=p(X,Z_{1})'q_{0}(V),
\]
where $p(X,Z_{1})$ are known functions of $(X,Z_{1}')'$,
and the CRF (\ref{eq:controlregression_2}) takes the form 
\[
E\left[Y\mid X,Z_{1},V\right]=p_{0}(X)'q(V,Z_{1}),
\]
where $q(Z_{1},V)$ are known functions of $(Z_{1}',V)'$.
\end{rem}

\subsection{Average Structural Function}

For both classes of models in Assumptions \ref{ass:Assumption1p} and \ref{ass:Assumption1q}, the integrated CRF $\int E[Y |X,V=v] dF_{V}(v)$ is linear in the treatment functions and, in general, need not coincide with the ASF $\int g(X,\varepsilon) dF_{\varepsilon}(\varepsilon)$. 
We thus restrict outcome functions to those with implied ASF that does coincide with the integrated CRF.  
In this way, both the CRF and the ASF are linear in the treatment functions, and the ASF is expressed as a known functional of observable or estimable quantities only. 
This relation will form the basis of our ASF identification strategy.
\begin{assumptionenv} \label{ass:DGP}
    For the outcome model $Y=g(X,\varepsilon)$, there exists a control variable $V$ such that the relation
    \begin{equation} \label{eq:model}
        \int g(X,\varepsilon) dF_{\varepsilon}(\varepsilon) = \varphi(X)'E[\chi(V)],
    \end{equation}
holds with either $(\varphi(X)',\chi(V)')'=(p(X)',q_{0}(V)')'$ under Assumption \ref{ass:Assumption1p}, or with $(\varphi(X)',\chi(V)')'=(p_{0}(X)',q(V)')'$ under Assumption \ref{ass:Assumption1q}, whichever is assumed.
\end{assumptionenv}

Assumption \ref{ass:DGP} encapsulates two types of restrictions. 
First, a functional form restriction on $g(X,\varepsilon)$, with implied ASF required to be in linear form. 
The ASF derivative is also in linear form, such that 
$\partial\mu(X)/\partial x=\{\partial \varphi(X)/\partial x\}^{\prime}E[\chi(V)]$, 
a derivative version of the average treatment effect. 
Assumption \ref{ass:DGP} is a generalization of the heterogeneous coefficients functional form 
(\ref{eq:hcoefs_model_p}),
and hence of the models in \citet{Flo Heck Meg Vytl 2008} and \citet{MT:2016}.\footnote{For an example of outcome function not linear in known functions $p(X)$ while the implied ASF is, consider 
$g(X,\varepsilon) = p(X)'\beta(\varepsilon) + \xi(X,\varepsilon^{*})$, 
with $\varepsilon^{*}\mapsto \xi(X,\varepsilon^{*})$ an odd function, 
and $\varepsilon^{*}$ a component of $\varepsilon$ with distribution symmetric about zero. Then $\int \xi(X,\varepsilon^{*}) dF_{\varepsilon^{*}}(\varepsilon^{*})=0$, and hence: 
$\int g(X,\varepsilon) dF_{\varepsilon}(\varepsilon)
=\int \{ p(X)'\beta(\varepsilon) + \xi(X,\varepsilon^{*}) \} dF_{\varepsilon}(\varepsilon) 
=p(X)'E[\beta(\varepsilon)] + \int \xi(X,\varepsilon^{*}) dF_{\varepsilon^{*}}(\varepsilon^{*})
=p(X)'E[\beta(\varepsilon)]$.} 
Assumption \ref{ass:DGP} also generalizes these models by allowing for 
discrete or continuous treatment with an observable control variable.
Moreover, under restrictions on how $V$ affects the CRF in Assumption \ref{ass:Assumption1q} and regularity conditions on $g(X,\varepsilon)$ in footnote \ref{fn:approximation}, our formulation allows for unrestricted $g(X,\varepsilon)$. 

Second, a restriction on the conditional distribution of $\varepsilon$ given $X$ and $V$, 
imposing a form of conditional independence between $\varepsilon$ and $X$ given $V$, 
with the ASF required to coincide with the integrated CRF. This is made apparent by writing (\ref{eq:model}) as
\begin{equation} \label{eq:implicit_independence}
\int g(X,\varepsilon) dF_{\varepsilon}(\varepsilon) = \int \left\{ \int g(X,\varepsilon) dF_{\varepsilon \mid X,V}(\varepsilon \mid X,v) \right\} dF_V(v),
\end{equation}
by $\varphi(X)' E[\chi(V)] = \int \{\varphi(X)'\chi(v)\}dF_V(v) = \int E[Y|X,V=v]dF_{V}(v)$ under Assumption \ref{ass:Assumption1p} or \ref{ass:Assumption1q}. 
Clearly, for general $g(X,\varepsilon)$ 
independence of $\varepsilon$ and $X$ given $V$ is sufficient for (\ref{eq:implicit_independence}), and hence also for (\ref{eq:model}) under Assumption \ref{ass:Assumption1p} or \ref{ass:Assumption1q}. 

For restricted $g(X,\varepsilon)$, (\ref{eq:model}) can hold under weaker forms of conditional independence. 
In the leading examples (\ref{eq:hcoefs_model_p}) 
and (\ref{eq:hcoefs_model_p*})-(\ref{eq:cond mean het coefs}), 
relation (\ref{eq:model}) holds under 
mean independence of $\beta(\varepsilon)$ and $X$ given $V$. 
In these examples, the ASF is $\mu(X)= p(X)^{\prime}E[\beta(\varepsilon)]$ and $\mu(X)= p^{*}(X)^{\prime}E[\beta(\varepsilon)]$, respectively; 
see \citet{Cham:1984} and \citet{Wool:2005}. 
Relation (\ref{eq:model}) is then verified by expressing the ASF as a 
linear combination of 
$E[q_{0}(V)]$ and $p_{0}(X)=\Omega p^{*}(X)$, respectively. 
For (\ref{eq:hcoefs_model_p}), by iterated expectations and $E[\beta(\varepsilon)| V]=q_{0}(V)$,
\begin{equation*} 
    \mu(X)
    =p(X)^{\prime}E[\beta(\varepsilon)]
    =p(X)^{\prime}E[E[\beta(\varepsilon)\mid V]]
    =p(X)^{\prime}E[q_{0}(V)].
\end{equation*}
Similarly, for 
(\ref{eq:hcoefs_model_p*})-(\ref{eq:cond mean het coefs}), by $E[\beta(\varepsilon)|V]=\Omega'q(V)$ and $\Omega p^{*}(X) = p_{0}(X)$,
\begin{equation*} 
    \mu(X)
    =p^{*}(X)^{\prime}E[\beta(\varepsilon)]
    =p^{*}(X)^{\prime}E[\Omega'q(V)]
    =\{\Omega p^{*}(X)\}'E[q(V)]
    =p_{0}(X)^{\prime}E[q(V)].
\end{equation*}

This discussion shows that Assumption \ref{ass:DGP} allows for a wide class of models, beyond heterogeneous coefficients models that are linear in known functions of the treatment.

\begin{rem}

ASF identification requires integrating over
the marginal distribution of 
$V$. There are
other interesting structural objects that do not rely only on the
marginal distribution of $V$. 
For example, using $\partial_{x}$ to denote partial derivatives with respect to $x$, 
in the leading examples (\ref{eq:hcoefs_model_p}) and (\ref{eq:hcoefs_model_p*}), with (\ref{eq:CMI}),  $g(X,\varepsilon)$ has average derivative 
\begin{eqnarray*}
E[\partial_{x}\{\varphi(X)^{\prime}\beta(\varepsilon)\}] & = & E[E[\partial_{x}\{\varphi(X)^{\prime}\beta(\varepsilon)\}|X,V]]=E[\{ \partial_{x} \varphi(X)\} ^{\prime}E[\beta(\varepsilon)|X,V]]\\
 & = & E[\{ \partial_{x} \varphi(X)\} ^{\prime}E[\beta(\varepsilon)\mid V]]=E[\partial_{x} E[Y\mid X,V]],
\end{eqnarray*}
as shown in \citet[p.1491]{Imbens Newey 2009} for general nonseparable models with $\varepsilon \ind X | V$. 
This object and others like it, including the local average response
of \citet{Altonji Matzkin 2005}, do not require full support for identification in general nonseparable models. For this reason, we focus our analysis on the ASF where we use the CRF's varying coefficients structure to weaken the full support
condition.

\end{rem}

\subsection{Triangular models}

An important kind of control variable arises in a triangular model
where an instrumental variable $Z$ is excluded from the outcome equation 
$Y=g(X,\varepsilon)$ and where $X$ is a scalar with 
\begin{equation} \label{eq:h(z,eta)}
X=h(Z,\eta),
\end{equation}
with $\eta\mapsto h(Z,\eta)$ strictly monotonic. If $(\varepsilon,\eta)$
is jointly independent of $Z$, 
then $\varepsilon$ is independent of $X$ given $V$ 
for $V=F_{X|Z}(X|Z)$, the CDF of $X$ conditional on $Z$ (\citet{Imbens Newey 2009}).\footnote{When strict monotonicity does not hold, for instance with discrete $X$, the control function is set-valued and structural objects of interest such as average and quantile treatment effects are partially identified. This case is considered in \citet{Chesher 2005} and \citet{HK2024}.} 
Alternatively, $V=F_{X|Z}(X|Z)$ is a control variable in the leading examples (\ref{eq:hcoefs_model_p}) and (\ref{eq:hcoefs_model_p*}) where the outcome
functions are in heterogeneous coefficients form,  
under the weaker conditions that $\eta$ is independent from $Z$
and that $\beta(\varepsilon)$ is mean independent of $Z$ conditional on
$\eta$.
\begin{thm} \label{thm:Theorem1}
For triangular models of the form (\ref{eq:hcoefs_model_p}) or  (\ref{eq:hcoefs_model_p*}), 
with (\ref{eq:h(z,eta)}), 
if $\eta$ is independent from $Z$ and $E[\beta(\varepsilon)|\eta,Z]=E[\beta(\varepsilon)|\eta]$,
then $E[\beta(\varepsilon)|X,V]=E[\beta(\varepsilon)|V]$.
\end{thm}

Let $x \mapsto h^{-1}(Z,x)$ denote the inverse function of $\eta \mapsto h(Z,\eta)$. 
Since $\eta=h^{-1}(Z,X)$ and $V=F_{X|Z}(X|Z)=F_\eta(h^{-1}(Z,X))$, the two leading examples in triangular form, i.e., (\ref{eq:hcoefs_model_p}) and (\ref{eq:hcoefs_model_p*}) augmented with (\ref{eq:h(z,eta)}), are particular cases of the class of nonseparable triangular models of the form:
\[ 
        Y=g(X,\varepsilon), \quad X=h(Z,\eta), \quad \eta\mid Z \sim F_{\eta}, \quad \int g(X,\varepsilon) dF_{\varepsilon}(\varepsilon) = \varphi(X)'E[\chi(F_\eta(\eta))],
\]
with $F_\eta$ the CDF of $\eta$, $E[Y|X,\eta]=\varphi(X)'\chi(F_\eta(\eta))$, and where either $\varphi(\cdot)$ or $\chi(\cdot)$ is known, but not both. 
This formulation characterizes the set of nonseparable triangular models to which the results in this paper will apply.

\section{Identification} \label{sec:Section3}

One main contribution of this paper is to highlight and show that
in our modeling framework the ASF is identified under nonsingularity of the second moment matrix of the known functions of either the treatment given the controls, or the controls given the treatment.

\begin{assumptionenv}\label{ass:cond_nonsingularity}
\hypertarget{ass:cond_nonsingularity_p}{(\textnormal{p})} $E[\{ p(X)'q_{0}(V) \} ^{2}]<\infty$ and $E[p(X)p(X)'|V]$ is nonsingular with probability one; \hypertarget{ass:cond_nonsingularity_q}{(\textnormal{q})} $E[\{ p_{0}(X)'q(V) \} ^{2}]<\infty$ and $E[q(V)q(V)'|X]$ is nonsingular with probability one.
\end{assumptionenv}

Assumption \refp{ass:cond_nonsingularity} 
is sufficient for identification
of the unknown coefficients $q_{0}(V)$ in CRF models with known $p(X)$, and Assumption \refq{ass:cond_nonsingularity}
is sufficient for identification of the unknown coefficients $p_{0}(X)$ in CRF models with known $q(V)$.
\begin{thm}
(i) Under Assumption \ref{ass:Assumption1p}, if Assumption \refp{ass:cond_nonsingularity} holds then $q_{0}(V)$ is identified. 
(ii) Under Assumption \ref{ass:Assumption1q}, if Assumption \refq{ass:cond_nonsingularity}
holds then $p_{0}(X)$ is identified.\label{thm:Theorem2}
\end{thm}
In Section \ref{sec:Section5} we discuss conditions under which $E[p(X)p(X)'|V]$
and $E[q(V)q(V)'|X]$ are nonsingular in triangular models.  
All those conditions are sufficient for
identification of $q_{0}(V)$ and of $p_{0}(X)$, respectively,
including those that allow for discrete valued instrumental variables. 
We also note that identification of $q_{0}(V)$
and of $p_{0}(X)$ means uniqueness on sets of $V$ and of $X$ having
probability one, respectively. Thus, under Assumption \ref{ass:DGP} with Assumption \ref{ass:Assumption1p}, the ASF will be identified as
\[
\mu(X)=p(X)'E[q_{0}(V)], 
\]
and, under Assumption \ref{ass:DGP} with Assumption  \ref{ass:Assumption1q}, as
\[
\mu(X)=p_{0}(X)'E[q(V)]. 
\]
In other words, the ASF is identified under Assumption \ref{ass:Assumption1p}
because $p\left(X\right)$ is a known function, and $q_{0}(V)$ is
identified, and hence $E[q_{0}(V)]$ also is; the ASF is identified
under Assumption \ref{ass:Assumption1q} because $p_{0}(X)$ is identified,
and $q(V)$ is a known function, and hence $E[q(V)]$ also is.
\begin{thm}
Suppose Assumption \ref{ass:DGP} holds. 
(i) Under Assumption \ref{ass:Assumption1p}, if Assumption \refp{ass:cond_nonsingularity} holds then the ASF is identified. 
(ii) Under Assumption \ref{ass:Assumption1q}, if Assumption \refq{ass:cond_nonsingularity} holds then the
ASF is identified. \label{thm:Theorem3}
\end{thm}
\begin{sloppy}
In CRF models with known $p(X)$, we use linearity of the ASF $p(X)'E[q_{0}(V)]$ in $p(X)$ to show that, for a sufficiently rich set of $p(X)$ values, nonsingularity
of $E[p(X)p(X)'|V]$ with probability one is also necessary for identification.
\par\end{sloppy}
\begin{thm} \label{thm:Theorem4}
Suppose Assumption \ref{ass:DGP} holds with Assumption \ref{ass:Assumption1p}, and $E[p(X)p(X)']$ is nonsingular. 
Then: if $E[p(X)p(X)'|V]$ is singular with positive probability then the ASF is not identified.
\end{thm}
In CRF models with known $q(V)$, the ASF $p_{0}(X)'E[q(V)]$ 
is now a linear combination of unknown functions $p_{0}(X)$ with known coefficients $E[q(V)]$.
In general, nonsingularity of $E[q(V)q(V)'|X]$ with probability one
is not necessary for ASF identification in this case. 
A condition that is both necessary and sufficient is that $E[q(V)]$ belongs to the range $\mathcal{R}(E[q(V)q(V)'|X])$ of $E[q(V)q(V)'|X]$ with probability one.
\begin{thm} \label{thm:Theorem5}
Suppose Assumption \ref{ass:DGP} holds with Assumption \ref{ass:Assumption1q}. Then: $E[q(V)]\in\mathcal{R}(E[q(V)q(V)'|X])$ with probability one if, and only if, the ASF is identified.
\end{thm}
When $E[q(V)q(V)'|X]$ is nonsingular with probability one, its range is $\mathbb{R}^{K}$ and hence the range condition is automatically satisfied. 
When conditional nonsingularity only holds with positive probability, then the range condition allows for identification of the ASF, but is in general
not sufficient for point identification of $p_{0}(X)$. Like 
nonsingularity, the range condition depends on observable or estimable quantities only.

\begin{rem} \label{rem:control CQF and CDF}
Conditional nonsingularity is also sufficient for other control regressions, such as control quantile regressions 
$Q_{Y|XV}(u|X,V)=p(X)^{\prime}q_{u}(V)$ or $p_{u}(X)^{\prime}q(V)$, $u\in(0,1)$, with $Y$ continuous and unknown coefficients $(q_{u}(V)', p_{u}(X)')'$, 
and control distribution regressions 
$F_{Y|XV}(y|X,V)=\Gamma(p(X)^{\prime}q_{y}(V))$
or $\Gamma(p_{y}(X)^{\prime}q(V))$, for $y\in \mathcal{Y}$ the support of $Y$, and with unknown coefficients $(q_{y}(V)', p_{y}(X)')'$ and $\Gamma$ a specified strictly increasing continuous CDF. 
For outcome models (\ref{eq:g(x,e)}) with implied control quantile and distribution regression functions of the forms given here, identification of the unknown coefficients implies identification of quantile and distributional treatment effects. These effects are known functionals of the control regressions above, which are identified when the coefficients are; cf. \citet[Section 4]{NeweyStouli2021} for a detailed exposition. 
These specifications generalize those in \citet{CFNSV:2020} to allow for unknown functions of $V$ or $X$. 
Section \ref{subsec:Alternative-model-specifications} applies our identification results to two nonseparable triangular models with bivariate unobserved 
heterogeneity in the outcome equation. 
\end{rem}

\section{Discussion of conditional nonsingularity} \label{sec:Section4}

\subsection{Conditional nonsingularity with probability one}

For CRF models with known $p(X)$, nonsingularity of $E[p(X)p(X)^{\prime}|V]$ allows for the conditional support of $X$ given $V$ to be a strict subset of the marginal support of $X$ with positive probability, for instance when the conditional support of $X$ given $V$ is discrete whereas both $X$ and $V$ are continuous. 
Under conditional nonsingularity, $p(X)$ known and identification of $q_{0}(V)$ together imply uniqueness of the CRF $p(X)'q_0(V)$ on a set of $(X,V)$ having probability one. Therefore the integral in characterization (\ref{eq:integrated_CRF}) of the ASF is well-defined for $E[Y|X,V]=p(X)'q_0(V)$ because integration then occurs over a range of $v$ values conditional on $X$ where the CRF is identified. 
Stronger sufficient conditions for identification are full support (\citet{Imbens Newey 2009}) and measurable separability (\citet{Flo Heck Meg Vytl 2008}), that any function of $X$ equal to a function of $V$ with probability one must be equal to a constant with probability one. Both conditions require $X$ to have continuous support conditional on $V$.

Conditional nonsingularity for identification in models of fixed dimension and full support for nonparametric identification are related under regularity conditions. 
For $p^{J}(X)$ denoting an increasing sequence of mean-square spanning approximating functions, nonsingularity for each element of this sequence implies full support, and hence also nonparametric identification. 
To state this result formally, denote the smallest and largest eigenvalue of a matrix $A$ by $\lambda_{\min}(A)$ and $\lambda_{\max}(A)$, respectively, and let $p^{J}(X)$ be mean-square spanning if, for any $f(X)$ such that $E[f(X)^{2}]<\infty$, there is $\gamma^{J}$ such that $E[\{f(X)-p^{J}(X)'\gamma^{J}\}^{2}]\rightarrow0$ as $J\rightarrow\infty$.
\begin{thm}
Suppose that $p^{J}(X)$ is mean-square spanning and, for all $J$,
we have $\lambda_{\min}(E[p^{J}(X)p^{J}(X)'|V])>0$ with probability
one and $\lambda_{\max}(E[p^{J}(X)p^{J}(X)'])\leq C$ for some finite
$C$. Then: for any set $\mathcal{A}$, if $\Pr(X\in\mathcal{A})>0$,
then $\Pr(X\in\mathcal{A}|V)>0$ with probability one.\label{thm:Theorem6}
\end{thm}
Theorem \ref{thm:Theorem6} is a novel kind of result that relates
semi- and non-parametric identification conditions, embedding
nonparametric identification as a particular case in a general class of identification results for flexible outcome models with ASF and CRF in linear form, under Assumption \ref{ass:DGP} with Assumption \ref{ass:Assumption1p}. 
Denoting by $q_{0}^{J}(V)$ the coefficients in CRFs with known functions $p^{J}(X)$, this result reveals that a unified treatment of semi- and non-parametric identification is achieved under conditional nonsingularity: 
by $p^{J}(X)$ known, nonsingularity for given $J$ implies 
identification of $p^{J}(X)'q_0^{J}(V)$ viewed as a correct CRF model, 
and nonsingularity for all $J$ implies uniqueness on a set of $(X,V)$ with probability one of each $p^{J}(X)'q_0^{J}(V)$ in a sequence of CRF approximating models, and hence is sufficient for nonparametric identification. 
This embedding of semi- and non-parametric identification into a common framework 
also applies to the other control regressions in Remark \ref{rem:control CQF and CDF}, with nonsingularity for all approximating models implying full support,  and hence also nonparametric identification of distributional and quantile treatment effects (\citet{Imbens Newey 2009}).

For the class of models defined by Assumption \ref{ass:DGP} with Assumption \ref{ass:Assumption1p}, i.e., with $p(X)$ known, a converse result is that full support implies conditional nonsingularity with probability one, under the additional maintained assumption that $E[p(X)p(X)']$ be nonsingular. 
This is a corollary of Theorem \ref{thm:Theorem4}.

\begin{cor} \label{cor:Corollary1}
Suppose 
Assumption \ref{ass:DGP} holds with Assumption \ref{ass:Assumption1p}, and $E[p(X)p(X)']$ is nonsingular. Then: with probability one, if the conditional support of $X$ given $V$ is the same as the marginal support of $X$, then $E[p(X)p(X)'|V]$ is nonsingular. 
\end{cor}

The relaxation of full support afforded by our framework is important in practice. We discuss below the leading case of triangular models with discrete instruments, where full support cannot hold.
More generally, within the class of models we consider, treatment effects  can be identified for continuous
$X$ when the joint support of $X$
and $V$ is not rectangular. For discrete $X$, treatment effects can be identified 
when the conditional support of $X$ given $V$ has fewer points than the marginal support of $X$ with positive probability, with $p(X)$ of dimension smaller than the number of treatment regimes.
\begin{rem}
A result analogous to Theorem \ref{thm:Theorem6} holds for mean-square spanning functions $q^{K}(V)$ assuming that, for all $K$, $\lambda_{\min}(E[q^{K}(V)q^{K}(V)'|X])>0$ with probability one and $\lambda_{\max}(E[q^{K}(V)q^{K}(V)'])\leq C$
for some finite $C$. The result is now that, for any set $\mathcal{A}$,
$\Pr(V\in\mathcal{A})>0$ implies $\Pr(V\in\mathcal{A}|X)>0$ with probability one. 
\begin{rem}
For mutually exclusive discrete treatments and $X$ a vector of dummy variables as in (\ref{eq:p(X)}), full support is the same as nonsingularity of $E[p(X)p(X)'|V]$ with probability one (\citet{NeweyStouli:2022}, Theorem 3, p. 868). 
Similarly, for the example in Section \ref{subsec:CRF_knownq} where $V$ is discrete and defined as a vector of dummies for mutually exclusive control regimes, nonsingularity of $E[q(V)q(V)'|X]$ with
probability one is the same as full support, and hence our results imply 
nonparametric identification of average treatment effects in that case.
\end{rem}
\end{rem}

\subsection{Conditional nonsingularity with positive probability}\label{subsec:CNwpp}

For CRF models with known $p(X)$, a condition weaker than conditional nonsingularity with probability one is nonsingularity of $E[p(X)p(X)'|V]$
with positive probability. This condition  has been used by \citet{MT:2016}
for identification in the triangular model formed by (\ref{eq:hcoefs_model_p}) 
and (\ref{eq:h(z,eta)}). Our identification results that are based on the control variable $V=F_{X|Z}(X|Z)$ are thus related to their approach. Suppose nonsingularity
of $E[p(X)p(X)'|V]$ holds on a set with positive probability, and $\overline{q}(v)\neq q_{0}(v)$ for a value $v$ in that set.
Then, with $\lambda(v)\equiv\overline{q}(v)-q_{0}(v)$, 
\begin{equation}
E[\{p(X)'\lambda(V)\}^{2}\mid V=v]=\lambda(v)^{\prime}E[p(X)p(X)'\mid V=v]\lambda(v)>0.\label{eq:id_v}
\end{equation}
If $E[\beta(\varepsilon)]$
and $E[p(X)p(X)']$ exist then the expectation in (\ref{eq:id_v})
exists, and hence $q_{0}(V)$ is identified from $E[Y|X,V=v]$, by definition (\ref{eq:controlregression_1}).
\citet{MT:2016} showed this, 
and noted that $E[q_{0}(V)]$ is identified if the set of $v$ with $E[p(X)p(X)'|V=v]$ 
nonsingular has probability one. Their approach is local (pointwise
in $v$) and constructive for $q_{0}(v)$. In contrast, directly considering
uniqueness with probability one of $q_{0}(V)$ is useful for our analysis, which focuses on  average treatment effects and hence requires identification of $E[q_{0}(V)]$, and is constructive for $q_{0}(V)$.

\begin{sloppy}
Compared to ASF identification in CRF models with known $p(X)$, nonsingularity of $E[q(V)q(V)'|X]$ on a set with positive probability is sufficient for ASF identification on that set in CRF models with known $q(V)$.

\begin{thm} \label{thm:Theorem17}
Suppose Assumption \ref{ass:DGP} holds with \ref{ass:Assumption1q}. If $E[q(V)q(V)'|X]$ is nonsingular on a set with positive probability, then the ASF is identified on that set.
\end{thm}
In contrast with the case where $p(X)$ is known, this result demonstrates that local features of the ASF can be identified without conditional nonsingularity holding with probability one, when restrictions are placed on how control variables affect the CRF, while allowing for the general nonseparable outcome model in  (\ref{eq:g(x,e)}).

Identification conditions based on nonsingularity of either $E[p(X)p(X)'|V]$ or $E[q(V)q(V)'|X]$ with positive probability have been used by \citet{NeweyStouli2021} for parametric CRFs of the form
\begin{equation}
E[Y\mid X,V]=w(X,V)'\beta,\quad w(X,V)\equiv p(X)\otimes q(V),\label{eq:paramCRF}
\end{equation}
where now both $p(X)$ and $q(V)$ are known functions, and with $\otimes$ denoting the kronecker product. This CRF has either the varying coefficient structure (\ref{eq:controlregression_1}),
\[
w(X,V)'\beta=\sum_{j=1}^{J}p_{j}(X)'\{q(V)'\beta_{j}\}=p(X)'q_{0}(V),\quad\beta=(\beta_{1}',\ldots,\beta_{J}')',
\]
with $q_{0}(V)=(q_{01}(V),\ldots,q_{0J}(V))'$ and $q_{0j}(V)\equiv q(V)'\beta_{j}$,
$j\in\{1,\ldots,J\}$, or the varying coefficient structure (\ref{eq:controlregression_2}),
\[
w(X,V)'\beta=\sum_{k=1}^{K}q_{k}(V)'\{p(X)'\beta_{k}\}=p_{0}(X)'q(V),\quad\beta=(\beta_{1}',\ldots,\beta_{K}')',
\]
with $p_{0}(X)=(p_{01}(X),\ldots,p_{0K}(X))'$ and $p_{0k}(X)=p(X)'\beta_{k}$,
$k\in\{1,\ldots,K\}$, and is thus related to our 
modeling framework. By standard results such as those of \citet{NMcF:1994},
identification of $\beta$ requires nonsingularity of $E[w(X,V)w(X,V)']$,
for which \citet{NeweyStouli2021} give sufficient conditions when
the smallest eigenvalue of either $E[p(X)p(X)'|V]$ or $E[q(V)q(V)'|X]$
is bounded away from zero uniformly over a set of $V$ or $X$, respectively,
with positive probability.

Parametric models with CRFs of the form (\ref{eq:paramCRF}) are particular
cases of (\ref{eq:controlregression_1}) and of (\ref{eq:controlregression_2}), with both $p(X)$ and $q(V)$
known. For example, an important particular case is CRF models with multiple treatments and interaction terms proposed by \citet[Section 3.1.1, p.77]{NeweyStouli2021}, formed by setting $p(X)=(1,X(1),\ldots,X(T))'$ as in (\ref{eq:p(X)}), and by taking $q(V)=(1,\widetilde{q}(V)')'$ for $\widetilde{q}(V)$ a vector of centered known functions, i.e., 
with $E[\widetilde{q}(V)]=0$. With $\widetilde{q}(V)= V-E[V]$,\footnote{An alternative for triangular models where $V=F_{X|Z}(X|Z)$ is to use $\widetilde{q}(V)= \Phi^{-1}(V)$ (cf. \citet{CFNSV:2020}).} 
this specification gives
\[
E[Y\mid X,V]=
\beta_{1}'p(X)+\{\beta_{2}'p(X)\}\{V-E[V]\},
\]
which is of the form $w(X,V)'\beta$ in (\ref{eq:paramCRF}) with
$\beta=(\beta_{1}',\beta_{2}')'$. By \citet{NeweyStouli2021}, a
sufficient condition for identification of $\beta$ is that for a
set $\widetilde{\mathcal{V}}$ of values of $V$ with $\Pr(\widetilde{\mathcal{V}})>0$
such that $E[p(X)p(X)'|V]$ is nonsingular, the matrix $E[1(V\in\widetilde{\mathcal{V}})q(V)q(V)']$
is nonsingular. This condition can be much weaker than common support
when the number of points in the support of $V$ is greater than two. Equivalently, this shows that regression models with multiple treatments and interaction terms, 
of the form
\[
Y=\beta_{1}'p(X)+\{\beta_{2}'p(X)\}\widetilde{q}(V)+U,\quad E[U|X,V]=0,
\]
can identify average treatment effects without common support or strong overlap.\footnote{For this model, the results in \citet{FvVV24} apply for least-squares estimation and uniform inference on both the implied CRF and ASF; see also \citet[Section 4.2]{GPHK2022} for a discussion of this model for estimation of average treatment effects.} 
\par\end{sloppy}

\section{Identification in triangular models} \label{sec:Section5}

Nonsingularity of $E[p(X)p(X)'|V]$ and of $E[q(V)q(V)'|X]$ are generic
conditions for identification in 
model specifications above, for any observable or estimable $V$. These
conditions, however, do not use the specific structure of triangular
models. Explicitly accounting for their specific features 
leads to the formulation of primitive conditions that can
be considerably easier to interpret and verify. 
Thus we specialize our identification analysis to triangular models with implied CRF of the form (\ref{eq:controlregression_1}), 
with known functions of $X$. Results with known functions of $V$
can be derived using analogous arguments and are summarized in Remark
\ref{rem:model2} below.

In triangular models with control variable $V=F_{X|Z}(X|Z)$, the
identification conditions can equivalently be stated in terms of the
first-stage representation $X=Q_{X|Z}(V|Z)$ and the instrument $Z$.
By independence of $V$ from $Z$, the identification condition with
(\ref{eq:controlregression_1}) is that 
\[
E[p(Q_{X\mid Z}(v\mid Z))p(Q_{X\mid Z}(v\mid Z))'] 
\]
be nonsingular for almost every (a.e.) $v$ in the support $\mathcal{V}$
of $V$.

\begin{sloppy}
When $p(X)=(1,X)'$, the condition is that the second moment matrix of $(1,Q_{X|Z}(v|Z))'$ is nonsingular for a.e. $v\in\mathcal{V}$, which is the same as
\begin{equation}
\text{det}\left(E[p(Q_{X\mid Z}(v\mid Z))p(Q_{X\mid Z}(v\mid Z))']\right)=\text{Var}(Q_{X\mid Z}(v\mid Z))>0,\label{eq:ScalarCondition}
\end{equation}
for a.e. $v\in\mathcal{V}$. 

When $p(X)$ includes a vector $\widetilde{p}(X)$ of known transformations
of $X$, the identification condition is that the second moment matrix
of 
\[
p(Q_{X\mid Z}(v\mid Z))=(1,\widetilde{p}(Q_{X\mid Z}(v\mid Z))')'
\]
is nonsingular for a.e. $v\in\mathcal{V}$, which is the same as the
condition that the variance matrix of $\widetilde{p}(Q_{X\mid Z}(v|Z))$
is nonsingular for a.e. $v\in\mathcal{V}$.
\begin{thm}
$E[p(X)p(X)'|V]$ is nonsingular with probability one if, and only if, the variance
matrix $\text{Var}(\widetilde{p}(Q_{X\mid Z}(v|Z)))$ is nonsingular
for a.e. $v\in\mathcal{V}$.\label{thm:Theorem8}
\end{thm}
Theorem \ref{thm:Theorem8} relates identification in triangular models
to nonsingularity of a collection of variance matrices of the treatment functions, generalizing (\ref{eq:ScalarCondition}) to the case of vector $\widetilde{p}(X)$. This formulation exploits the implications of independence of $V$ from $Z$, providing a primitive characterization of nonsingularity for $E[p(X)p(X)^{\prime}|V]$. This formulation also helps clarify that measurable separability can fail while our condition holds. In their first example, \citet[p.1198]{Flo Heck Meg Vytl 2008}
consider a binary instrument $Z\in\{0,1\}$ and a first-stage model
of the form
\begin{equation}
X=Z+\eta,\label{eq:FS}
\end{equation}
where $\eta$ is uniformly distributed on the unit interval $[0,1]$.
They show that in this simple example $X$ and $\eta$ are not measurably
separable. In contrast, for model (\ref{eq:FS}) and with $V=F_{\eta}(\eta)$,
we have that $\textrm{Var}(Q_{X\mid Z}(v|Z))>0$ for a.e. $v\in(0,1)$.
Therefore, if in addition 
\[
Y=p(X)'\beta(\varepsilon),\quad p(X)=(1,X)',
\]
then $E[p(X)p(X)'|V]$ is nonsingular with probability one. This provides
a simple example of failure of measurable separability while our condition
holds, and establishes that measurable separability is not necessary
for identification.
\end{sloppy}

When $Z$ has discrete support $\mathcal{Z}=\left\{ z:\Pr(Z=z)\geq\delta>0\right\} $
of finite cardinality $|\mathcal{Z}|$, a necessary condition for
nonsingularity is that the set $\mathcal{Q}(V)$ of distinct values 
of $z\mapsto Q_{X\mid Z}(V|z)$ has cardinality $|\mathcal{Q}(V)|$
greater than or equal to $J=\dim(p(X))$ with probability one.\footnote{Formally, for $v\in(0,1)$, we define $\mathcal{Q}(v)=\left\{ Q_{X\mid Z}(v\mid z_{m})\right\} _{m\in\mathcal{M}(v)}$,
where 
\[
\mathcal{M}(v)=\left\{ m\in\{1,\ldots,|\mathcal{Z}|\}:Q_{X\mid Z}(v\mid z_{m})\neq Q_{X\mid Z}(v\mid z_{m'})\,\textrm{for all }m'\in\{1,\ldots,|\mathcal{Z}|\}\backslash\{m\}\right\} .
\]
}  Thus, for a sufficiently rich set of $p(X)$ values, Theorem \ref{thm:Theorem4} implies that the ASF cannot be identified if $|\mathcal{Z}|<J$.
\begin{thm}
Suppose $E[p(X)p(X)']$ is nonsingular. If $|\mathcal{Z}|<J$ then the ASF is not identified.\label{thm:Theorem9} 
\end{thm}
Theorem \ref{thm:Theorem9} formalizes the intuitive notion that the
complexity of the model, as measured by the dimension of its known 
component $p(X)$, is restricted by the cardinality of the
set of instrumental values: the ASF can only be identified when the number of support points in $\mathcal{Z}$ is not smaller than $J$, the number of treatment
functions. 
Thus only when $p(X)$ is two-dimensional can identification be achieved in the presence of a binary instrument $Z\in\{0,1\}$. 
A more primitive condition for identification in this case is that
a change in the value of the instrument shifts the value of the conditional
quantile function $z\mapsto Q_{X\mid Z}(V|z)$ with probability one,
the condition stated in \citet[p.1002]{MT:2016} that $\text{Var}(Q_{X\mid Z}(v|Z))>0$
for a.e. $v\in\mathcal{V}$. Here we further show that this condition is also necessary when $E[p(X)p(X)']$ is nonsingular. \citet{L:2024} gives a related analysis
in a panel random coefficient model.

\begin{rem}  \label{rem:model2}
For CRF model (\ref{eq:controlregression_2}) with known $q(V)$ and $V=F_{X|Z}(X|Z)$, a sufficient condition for identification is
that 
\[
E\left[q(F_{X\mid Z}(X\mid Z))q(F_{X\mid Z}(X\mid Z))'\mid X\right] 
\]
be nonsingular with probability one. In the simple case $q(V)=(1,V)'$,
the condition is that $\text{Var}(F_{X|Z}(X|Z)|X)>0$ with probability
one. 
With $|\mathcal{Z}| = 2$, the condition holds if, and only if, $F_{X\mid Z}(X|0)\neq F_{X\mid Z}(X|1)$ with probability one. When $q(V)=(1,\widetilde{q}(V)')'$ with $\widetilde{q}(V)$ a vector of known transformations of $V$, the condition is that $\text{Var}(\widetilde{q}(F_{X\mid Z}(X|Z))|X)$ be nonsingular with probability one, which can only hold for $|\mathcal{Z}| \geq K$. 
With continuous treatment, these conditions allow for identification in general nonseparable triangular models with binary or discrete instruments. 
Identification based on these conditions is constructive for series least-squares estimation of the unknown functions $p_{0}(X)$ (cf. Remark \ref{rem:estimation p_{0}}).

\end{rem}

\subsection{Alternative model specifications\label{subsec:Alternative-model-specifications}}

We present  two examples that illustrate implications of our identification
analysis for quantile and distributional treatment effects in alternative
control quantile and distribution regression model specifications.
To the best of our knowledge, identification with discrete instruments
has not been previously established in these models.
\begin{example} \label{Example1}
With $Y$ continuous, consider the recursive triangular model, 
\begin{equation}
Y=p(X)'\beta(\varepsilon,\eta),\quad X=h(Z,\eta),\quad(\varepsilon,\eta)\ind Z,\label{eq:BivariateCQR}
\end{equation}
where $\eta\mapsto h(Z,\eta)$ and $\varepsilon\mapsto p(X)'\beta(\varepsilon,\eta)$
are strictly increasing with probability one. With $V=F_{X|Z}(X|Z)$,
the corresponding control quantile regression function is
\[
Q_{Y\mid XV}(u\mid X,V)=p(X)'q_{u}(V),\quad q_{u}(V)\equiv\beta\left(Q_{\varepsilon\mid\eta}(u\mid Q_{\eta}(V)),Q_{\eta}(V)\right), \quad u\in(0,1).
\]
With $p(X)=(1,\widetilde{p}(X)')'$ known, we have that $q_{u}(V)$, and hence also $p(X)'q_{u}(V)$,
is identified for each $u\in(0,1)$ if $\text{Var}(\widetilde{p}(Q_{X\mid Z}(v|Z)))$ is nonsingular for a.e. $v\in\mathcal{V}$, by Theorem \ref{thm:Theorem8} and repeated application of Lemma \ref{lem:Lemma1} (cf. proof of Theorem \ref{thm:Theorem2}) with $m(X,V)=Q_{Y|XV}(u|X,V)$, for each $u\in (0,1)$.  
Here, $q_{u}(V)$ is a vector of bivariate heterogeneous coefficients that capture the effect of $\widetilde{p}(X)$ across the joint distribution of $(\varepsilon,\eta)$. 
Since quantile and distributional treatment effects are known functionals of $F_{Y\mid XV}(Y|X,V)$ (\citet[p. 1489]{Imbens Newey 2009}),
and $F_{Y\mid XV}(Y|X,V)=\int_{0}^{1}1(Q_{Y|XV}(u|X,V)\leq Y)du$ with $Q_{Y|XV}(u|X,V)=p(X)'q_{u}(V)$, these effects are identified in model (\ref{eq:BivariateCQR}).

\begin{sloppy}
This model generalizes the quantile specifications considered in \citet{MaKo:2006},
\citet{Lee:2007}, \citet{Jun:2009},\footnote{For the particular case $p(X)=(1,X)'$, \citet{Jun:2009} considers
a control quantile regression specification of the form $Q_{Y|XV}(u|X,V)=p(X)'q_{u}(V)$, $u\in(0,1)$, with first-stage equation $X=\pi_{0}(V)+Z'\pi_{1}(V)$,
and shows identification of $q_{u}(v)$ assuming that $E[ZZ']$ is
nonsingular. Our results show that neither the restrictions on the
first-stage functional form nor those on $p(X)$ are necessary for
identification of $q_{u}(V)$ with discrete $Z$.} \citet{Chern Fern Kow}, and \citet{CFNSV:2020}.\footnote{The conditions in these papers do not simultaneously allow for $p(X)$
to be any known function of $X$, $z\mapsto h(z,\eta)$ to be unknown,
$Z$ to be discrete, and the parameters $\beta(\varepsilon,\eta)$
to be bivariate.} This model is  a particular case of the recursive models in \citet{Chesher 2003}
and \citet{Imbens Newey 2009}, who do not allow for discrete instruments,
and of \citet{Chesher 2007}.\footnote{\citet{Chesher 2007} considers identification of partial differences
of the outcome function at particular values of the treatment and
$(\varepsilon,\eta)$ that depend on the distribution of $(X,Z,\varepsilon)$
under local quantile independence conditions.} Because $\eta$ is allowed to enter the outcome equation, this model
is not a particular case of those considered by \citet{Fev Hault 2015}
and \citet{Torgo 2015}, and their results do not apply here.
\par\end{sloppy}

\end{example}

\begin{example} \label{exa:Example2}
Let $\Gamma_{\xi}$ and $\Gamma_{\eta}$ be some
specified strictly increasing continuous CDFs.  
With $Y$ continuous and $r(\eta)=(1,\widetilde{r}(\eta)')'$ for known functions $\widetilde{r}(\eta)$, consider the latent random coefficients model
\begin{equation} \label{eq:controlDR}
\xi=r(\eta)'\varepsilon,\quad\xi\mid X,\eta\sim\Gamma_{\xi},\quad\eta=h(Z,X),\quad\eta\mid Z\sim\Gamma_{\eta},\quad(\varepsilon,\eta)\ind Z,
\end{equation}
with $\varepsilon=\beta(Y,X)$, and both $x\mapsto h(Z,x)$
and $y\mapsto r(\eta)'\beta(y,X)$  strictly increasing with probability
one. This model is a generalization of the control distribution regression 
example in \citet[pp. 511-512]{CFNSV:2020} that restricts $x\mapsto\beta(Y,x)$
to be linear, and sets $r(\eta)=(1,\eta)'$ and $\Gamma_{\eta}=\Phi$.
For $V=F_{X\mid Z}(X|Z)$ and $\widetilde{q}(V)\equiv\widetilde{r}(\Gamma_{\eta}^{-1}(V))$,
the control distribution regression function corresponding to model
(\ref{eq:controlDR}) is, for $y\in \mathcal{Y}$,
\[
F_{Y\mid XV}(y\mid X,V)=\Gamma_{\xi}(p_{y}(X)'q(V)),\quad p_{y}(X)\equiv\beta(y,X),\quad q(V)=(1,\widetilde{q}(V))'.
\]
By $\widetilde{r}(\eta)$ and $\Gamma_{\eta}$ known, $q(V)=(1,\widetilde{q}(V)')'$ is also known. Thus, 
$p_{y}(X)$, and hence also 
$\Gamma_{\xi}(p_{y}(X)'q(V))$, 
is identified for each $y\in \mathcal{Y}$ if $\text{Var}(\widetilde{q}(F_{X\mid Z}(X|Z))|X)$
is nonsingular with probability one, 
by Remark \ref{rem:model2} and repeated application of Lemma \ref{lem:Lemma1} (cf. proof of Theorem \ref{thm:Theorem2}) with $m(X,V)=F_{Y|XV}(y|X,V)$, for each $y\in \mathcal{Y}$. 
Here, $p_{y}(X)$ is a vector
of bivariate heterogeneous coefficients that capture the effect of
$q(V)$ across the joint distribution of $(Y,X)$. 
Similarly to Example \ref{Example1}, identification of $\Gamma_{\xi}(p_{y}(X)'q(V))$ 
implies identification of quantile and distributional treatment effects in model (\ref{eq:controlDR}).
\end{example}

\section{Model testability} \label{sec:Section6}

This paper uses functional formal restrictions to achieve identification without full support. In this section we establish testability of the model specifications implied by these restrictions, and we characterize all testable implications. 

We first consider testability of the CRF model (\ref{eq:controlregression_1}) with known functions of $X$. 
Results for CRFs with known functions of $V$ are stated in Remark \ref{rem:Testability} below. 
\begin{thm}
Suppose $E[p(X)p(X)'|V]$ is nonsingular with probability one and $E[Y^{2}]<\infty$.
Then: $E[Y|X,V]=p(X)'q_{0}(V)$ for some $q_{0}(V)$ if, and only
if, for $q^{*}(V)\equiv E[p(X)p(X)'|V]^{-1}E[p(X)Y|V]$, we have $E[\{Y-p(X)'q^{*}(V)\}a(X,V)]=0$
for all $a(X,V)$ with $E[a(X,V)^{2}]<\infty$.\label{thm:Theorem10}
\end{thm}
Theorem \ref{thm:Theorem10} characterizes the complete set of orthogonality
conditions implied by 
CRF specification (\ref{eq:controlregression_1}), in terms of known
functions and observable or estimable random variables only. In particular
$q^{*}(V)$ is a vector of conditional least-squares projections of
$Y$ on $p(X)$ given $V$, where $Y$, $X$ and $V$ are each observable
or estimable.

There are several possible choices of test functions $a(X,V)$. One
natural approach to exploit this characterization of testable implications
of the model is to specify test functions $a(X,V)$ to be a vector of $L$ power
transformations of the CRF:
\begin{equation}
a(X,V)=(\{p(X)'q^{*}(V)\}^{2},\ldots,\{p(X)'q^{*}(V)\}^{L})'.\label{eq:Overid3}
\end{equation}
Orthogonality conditions of the form
\[
E[\{Y-p(X)'q^{*}(V)\}a(X,V)]=0
\]
with (\ref{eq:Overid3}) extend the classical approach of \citet{Ramsay:1969}
for specification testing of mean regression functions to the control
regression setting. Alternative choices of $a(X,V)$
are revealing functions of \citet{Bierens 1982} and \citet{SW:1998}.

A key implication of Theorem \ref{thm:Theorem10} is that if $E[Y|X,V]$ misspecified, i.e., is not of the specified form $p(X)'q_{0}(V)$, then $E[\{Y-p(X)'q^{*}(V)\}a(X,V)]\neq0$
for some test function $a(X,V)$ with $E[a(X,V)^{2}]<\infty$. 
This provides the basis of a test that can detect violations of the model specification. 
Empirical likelihood-based testing procedures for unconditional orthogonality conditions have been developed in a parametric setting, with $q^{*}(V)=q(V;\theta^{*})$ (\citet{DIN 2003}). 
It is beyond the scope of this paper to extend these approaches to the case with infinite-dimensional parameters.

\begin{sloppy}
\begin{rem} \label{rem:Testability}
Analogous testable implications can be characterized
for CRF model (\ref{eq:controlregression_2}) with known functions of $V$.
In that case, we have that $E[Y|X,V]=p_{0}(X)'q(V)$ for some $p_{0}(X)$
if, and only if, for $p^{*}(X)\equiv E[q(V)q(V)'|X]^{-1}E[q(V)Y|X]$,
we have $E[\{Y-p^{*}(X)'q(V)\}a(X,V)]=0$ for all $a(X,V)$ with $E[a(X,V)^{2}]<\infty$. 
\end{rem}
\end{sloppy}

\section{Estimation\label{sec:Estimation}} \label{sec:Section7}

Our identification analysis leads to direct estimation methods for
the heterogeneous coefficients models we consider. One approach to
making estimation feasible is through approximation of the nonparametric
components $q_{0}(V)$ or $p_{0}(X)$ by approximating functions such
as splines or wavelets. Here we focus on $q_{0}(V)$ in CRF model (\ref{eq:controlregression_1}), and an estimator for $p_{0}(X)$
in CRF model (\ref{eq:controlregression_2}) can
be constructed analogously.

For the specification $E[Y|X,V]=p(X)'q_{0}(V)$, we approximate
each component $q_{0j}(V)$, $j\in\{1,\ldots,J\}$, of the unknown
functional coefficient vector $q_{0}(V)$ by a linear combination
of $K$ basis functions $\psi^{K}=(\psi_{1}^{K},\ldots,\psi_{K}^{K})'$,
\begin{equation}
q_{0j}(V)\approx\sum_{k=1}^{K}b_{jk}\psi_{k}^{K}(V)=b_{j}'\psi^{K}(V),\quad j\in\{1,\ldots,J\},\label{eq:Approximation}
\end{equation}
where $b_{j}=(b_{j1},\ldots,b_{jK})'$, which yields an approximation
of the form 
\[
E\left[Y\mid X,V\right]=p(X)'q_{0}(V)\approx\sum_{j=1}^{J}\left\{ b_{j}'\psi^{K}(V)\right\} p_{j}(X)=b'[p(X)\otimes\psi^{K}(V)],
\]
where $b=(b_{1}',\ldots,b_{J}')'$. Such an approximation is well-defined
under our conditions with $b=b_{\textrm{LS}}^{K}$, the coefficient
vector of a least squares regression of $Y$ on $p(X)\otimes\psi^{K}(V)$,
\begin{equation}
b_{\textrm{LS}}^{K}\equiv\arg\min_{b\in\mathbb{R}^{JK}}E[\{ Y-b'[p(X)\otimes\psi^{K}(V)]\} ^{2}].\label{eq:OLS1}
\end{equation}

The proposed approximation is valid for the CRF $E[Y|X,V]$ if the
specified basis functions satisfy the following condition.

\begin{assumptionenv} For all $K$, $E[||\psi^{K}(V)||^{2}]<\infty$,
$E[\psi^{K}(V)\psi^{K}(V)^{\prime}]$ exists and is nonsingular, and,
for any $J$ vector of functions $a(V)$ with $E[||a(V)||^{2}]<\infty$,
there are $K\times1$ vectors $\varphi_{j}^{K}$, $j\in\{1,\ldots,J\}$,
such that as $K\rightarrow\infty$, $E[\sum_{j=1}^{J}\{a_{j}(V)-\psi^{K}(V)^{\prime}\varphi_{j}^{K}\}^{2}]\rightarrow0$.\label{ass:Assumption3-sieve}
\end{assumptionenv}

\begin{sloppy}Under this assumption, $E[Y|X,V]$ can be approximated
arbitrarily well by increasing the number of terms in the approximate
specification (\ref{eq:Approximation}).
\begin{thm}
Suppose that Assumptions \ref{ass:Assumption1p}, 
\refp{ass:cond_nonsingularity},
and \ref{ass:Assumption3-sieve} hold, and that $\sup_{v\in\mathcal{V}}E[||p(X)||^{2}|V=v]\leq C$
for some finite $C$. Then, 
\[
E[\{ E[Y\mid X,V]-[p(X)\otimes\psi^{K}(V)]^{\prime}b_{\text{LS}}^{K}\} ^{2}]\rightarrow0,
\]
as $K\rightarrow\infty$.\label{thm:Theorem11} 
\end{thm}
An estimator for the CRF is given by taking the sample analog in (\ref{eq:OLS1}),
upon substituting for the control variable $V$ by its estimated version
when it is unobservable. The properties of the corresponding ASF estimator,
including convergence rates and asymptotic normality, have been extensively
analysed by \citet{Imbens Newey 2002} for the general case where both $p(X)$ and $q(V)$ are increasing sequences of splines or power series approximating functions and the vector of regressors is of the kronecker product form we consider (cf. Theorems 6--8 in \citet{Imbens Newey 2002}).
Their analysis accounts for a first step nonparametric estimate of
the control variable, and their results directly apply to the simpler
case we consider here where the dimension of $p(X)$ is fixed, including
when $V$ is observable. In particular, we find that the convergence
rate for the ASF in the model is solely determined by the rate of
the first step estimator for the control variable. An immediate and
remarkable corollary of this result is that average treatment effects
are estimable at a parametric rate when $V$ is itself estimable at
a parametric rate or observable.\footnote{Models that allow for estimation of the CDF $F_{X|Z}(X|Z)$ at a parametric rate can be formulated using quantile and distribution regression (\citet{CFNSV:2020}) or dual and Gaussian transform regression (\citet{SS:2018,SS:2020}).}

\end{sloppy}
\begin{rem}\label{rem:estimation p_{0}}
An estimator for $p_{0}(X)$ in CRF models with known $q(V)$ 
can be constructed analogously, upon approximating each component
of the unknown coefficient $p_{0}(X)$ by a linear combination of approximating basis functions, for a fixed vector of known functions of $V$. 
Because the ASF is now a linear combination of unknown functions $p_0(X)$, the convergence rate for the implied ASF estimator is determined by both the first
step estimator for the control variable and the second step estimator
for $p_{0}(X)$. 
The results of \citet{Imbens Newey 2002} apply to this estimator as well.
\begin{rem}
When $V$ is a high-dimensional vector of observable controls, recent
methods can also be used to estimate the ASF. The debiased machine
learning estimation and inference methods of \citet{SC:2021}
and \citet{K:2021} apply to our setting, allowing for high dimensional basis functions $\psi^{K}(V)$. 
 
\end{rem}
\end{rem}

\section{Conclusion} \label{sec:Section8}

This paper introduces a unified modeling framework for treatment effects under minimal identification conditions. 
This framework is general enough to encompass a wide range of models of interest to applied researchers, and we provide a comprehensive treatment of identification, model testability and estimation for all classes of models considered. 
For flexible models of increasing dimension, we elucidate and characterize the connection between conditional nonsingularity for identification in these models, and the full support condition for nonparametric identification of treatment effects. 
In the presence of multidimensional heterogeneity and discrete instruments, we give conditions for identification of average treatment effects in general nonseparable triangular models, and our results demonstrate testability of both identification and 
model specification. 
These results extend to other types of control regressions, and our models can be conveniently estimated by series-based least squares estimators with well-understood properties.

\appendix

\section{Proofs\label{sec:Proofs}}

\subsection{Proof of Theorem \ref{thm:Theorem1}}

As in the proof of Theorem 1 in \citet{Imbens Newey 2009}, $V$ is
a one-to-one function of $\eta$. Then by equation (\ref{eq:h(z,eta)}),
iterated expectations, and conditional mean independence, 
\begin{align*}
E[\beta(\varepsilon)|X,V]=E[\beta(\varepsilon)|h(Z,\eta),\eta] & =E[E[\beta(\varepsilon)|\eta,Z]|h(Z,\eta),\eta]\\
 & =E[E[\beta(\varepsilon)|\eta]|h(Z,\eta),\eta]=E[\beta(\varepsilon)|\eta]=E[\beta(\varepsilon)|V],
\end{align*}
as claimed. \qed

\subsection{Proof of Theorem \ref{thm:Theorem2}}

We first state a useful lemma.
\begin{lem}
Given a random vector $(A',B')'$ and a specified function $m(A,B)$ with $E[m(A,B)^2]<\infty$, we have: if $m(A,B)=r(A)'s_{0}(B)$ for some vectors of unknown coefficients $s_{0}(B)$ and of known functions $r(A)$ with $E[r(A)r(A)'|B]$ nonsingular with probability one, then $s_{0}(B)$ is identified from $m(A,B)$.\label{lem:Lemma1}
\end{lem}

\begin{sloppy}
Let $\mathcal{B}$ denote the support of $B$ and $\lambda_{\min}(B)$ denote the smallest eigenvalue
of $E[r(A)r(A)'|B]$. Suppose that $\bar{s}\left(B\right)\neq s_{0}\left(B\right)$
with positive probability on a set $\widetilde{\mathcal{B}} \subseteq \mathcal{B}$, and
note that $\lambda_{\min}(B)>0$ on $\mathcal{B}$ by assumption. 
Then 
\begin{eqnarray*}
E[\{r(A)'\{\bar{s}(B)-s_{0}(B)\}\}^{2}] & = & E[\{\bar{s}(B)-s_{0}(B)\}'E[r(A)r(A)'\mid B]\{\bar{s}(B)-s_{0}(B)\}]\\
 & \geq & E[\left\Vert \bar{s}(B)-s_{0}(B)\right\Vert ^{2}\lambda_{\min}(B)]\\
 & \geq & E[1(B\in\mathcal{B}\cap\widetilde{\mathcal{B}})\left\Vert \bar{s}(B)-s_{0}(B)\right\Vert ^{2}\lambda_{\min}(B)]
\end{eqnarray*}
By definition $\Pr(\widetilde{\mathcal{B}})>0$ and $\widetilde{\mathcal{B}}\subseteq\mathcal{B}$
so that $\widetilde{\mathcal{B}}\cap\mathcal{B}=\widetilde{\mathcal{B}}$.
Thus the fact that $\left\Vert \bar{s}\left(B\right)-s_{0}\left(B\right)\right\Vert ^{2}\lambda_{\min}\left(B\right)$
is positive on $\widetilde{\mathcal{B}}\cap\mathcal{B}$ implies 
\[
E[1(B\in\mathcal{B}\cap\widetilde{\mathcal{B}})\left\Vert \bar{s}(B)-s_{0}(B)\right\Vert ^{2}\lambda_{\min}(B)]>0.
\]
We have shown that, for $\bar{s}\left(B\right)\neq s_{0}\left(B\right)$
with positive probability on a set $\widetilde{\mathcal{B}}$, 
\[
E[\{ r(A)^{\prime}\{ \bar{s}(B)-s_{0}(B)\} \} ^{2}]>0,
\]
which implies $r(A)^{\prime}\bar{s}\left(B\right)\neq r(A)^{\prime}s_{0}\left(B\right)$. Hence $s_{0}\left(B\right)$ is identified from $m(A,B)$.\qed
\end{sloppy}

We now give the proof of Theorem \ref{thm:Theorem2}:

Let $m(A,B)=E[Y|A,B]$. For Part (i), the result follows by application of Lemma \ref{lem:Lemma1} upon setting $r(A)=p(A)$ and $s_{0}(B)=q_{0}(B)$, with $A=X$, $B=V$; for Part (ii), the result follows upon setting $r(A)=q(A)$ and $s_{0}(B)=p_{0}(B)$, with $A=V$, $B=X$.

\subsection{Proof of Theorem \ref{thm:Theorem3}}

Part (i). Under Assumption \ref{ass:Assumption1p},
if $E[p(X)p(X)'|V]$ is nonsingular with probability one then $q_{0}(V)$
is identified by Theorem \ref{thm:Theorem2}(i), and hence $E[q_{0}(V)]$
also is. By $p(X)$ being a known function, $\mu(X)=p(X)'E[q_{0}(V)]$
is identified.

Part (ii). Under Assumption \ref{ass:Assumption1q},
if $E[q(V)q(V)'|X]$ is nonsingular with probability one then $p_{0}(X)$
is identified by Theorem \ref{thm:Theorem2}(ii). By $E[q(V)]$ being
a known vector, $\mu(X)=p_{0}(X)'E[q(V)]$ is identified.\qed

\subsection{Proof of Theorem \ref{thm:Theorem4}}

We first state a useful lemma.
\begin{lem} \label{lem:Lemma2}
Given a random variable \( V \) with support \( \mathcal{V} \), let \( \Xi(v) \) be a finite \( J \times J \) matrix defined for each \( v \in \mathcal{V} \), with null space $\mathcal{N}(\Xi(V))$. If \( \Xi(V) \) is singular with positive probability, then there exists \( \overline{\Delta}(V) \in \mathcal{N}(\Xi(V)) \) with 
\( E[\overline{\Delta}(V)]<\infty \) and \( E[\overline{\Delta}(V)] \neq 0 \).
\end{lem}

\begin{proof}
Since \( \Xi(V) \) is singular with positive probability, there exists a non-zero vector \( \Delta(V) \in \mathcal{N}(\Xi(V)) \) on a set \( \widetilde{\mathcal{V}} \subseteq \mathcal{V} \) such that \( \Pr(\widetilde{\mathcal{V}}) > 0 \). This means that \( \Delta(V) \neq 0 \) on \( \widetilde{\mathcal{V}} \). 
For each \( j \in \{1, \dots, J\} \), define the sets
\[
\widetilde{\mathcal{V}}_{j} = \{ v \in \widetilde{\mathcal{V}} : \Delta_{j}(v) \neq 0 \}.
\]
By construction, we have $\cup_{j=1}^{J} \widetilde{\mathcal{V}}_{j} = \{v\in \widetilde{\mathcal{V}} : \Delta(v)\neq0 \} = \widetilde{\mathcal{V}}$. Hence
\[
\Pr(\widetilde{\mathcal{V}}) = \Pr(\cup_{j=1}^{J}\widetilde{\mathcal{V}}_{j}) \leq \Sigma_{j=1}^{J} \Pr(\widetilde{\mathcal{V}}_{j}).
\]
Since \( \Pr(\widetilde{\mathcal{V}}) > 0 \), there must exist at least one \( j^{*} \in \{1, \dots, J\} \) such that \( \Pr(\widetilde{\mathcal{V}}_{j^{*}}) > 0 \).

Next, set $\widetilde{\Delta}(v)=\Delta(v)$
for $v\in\widetilde{\mathcal{V}}_{j^{*}}$, and $\widetilde{\Delta}(v)=0$
otherwise. 
By construction, \( \widetilde{\Delta}(V) \in \mathcal{N}(\Xi(V)) \), and \( \widetilde{\Delta}_{j^{*}}(V) \neq 0 \) on the set \( \widetilde{\mathcal{V}}_{j^{*}} \). 
Letting
\[
\overline{\Delta}(V) = \text{sign}\{\widetilde{\Delta}_{j^{*}}(V)\} \frac{\widetilde{\Delta}(V)}{||\widetilde{\Delta}(V)||},
\]
by construction we have $\overline{\Delta}_{j^{*}}(V)>0$ on $\widetilde{\mathcal{V}}_{j^{*}}$ and \( ||\overline{\Delta}(V)|| = 1 \), and hence \( E[||\overline{\Delta}(V)||] < \infty \) and $E[\overline{\Delta}_{j^{*}}(V)]\neq0$. 
Therefore, \( E[\overline{\Delta}(V)] \) is finite and non-zero. Since \( \overline{\Delta}(V) \in \mathcal{N}(\Xi(V)) \), this completes the proof.
\end{proof}

We now give the proof of Theorem \ref{thm:Theorem4}:

ASF identification means that there is no observationally equivalent $\overline{q}(V)\neq q_{0}(V)$ with positive probability such that $p(X)'E[\overline{q}(V)]\neq p(X)'E[q_{0}(V)]$. Following the proof of Theorem 1 in \citet{NeweyStouli:2022} with
discrete $X$, we show that if $E[p(X)p(X)'|V]$ is singular with
positive probability, then there exists an observationally equivalent
$\overline{q}(V)\neq q_{0}(V)$ with positive probability such that
$E[\overline{q}(V)]\neq E[q_{0}(V)]$.

Let $\Xi(V)= E[p(X)p(X)'|V]$. Then Lemma \ref{lem:Lemma2} implies that there exists $\overline{\Delta}(V) \in \mathcal{N}(E[p(X)p(X)'|V])$ such that $E[\overline{\Delta}(V)]<\infty$ and $E[\overline{\Delta}(V)] \neq 0$. 
Hence, upon setting $\overline{q}(V)\equiv q_{0}(V)+\overline{\Delta}(V)$, there exists an observationally equivalent $\overline{q}(V) \neq q_{0}(V)$ with $E[\overline{q}(V)]\neq E[q_{0}(V)]$. 
By nonsingularity of $E[p(X)p(X)']$, $p(X)'E[\overline{q}(V)]\neq p(X)'E[q_{0}(V)]$ and hence the ASF is not identified. \qed

\subsection{Proof of Theorem \ref{thm:Theorem5}}

ASF identification means that there is no observationally equivalent $\overline{p}(X)\neq p_{0}(X)$ with positive probability such that $\overline{p}(X)'E[q(V)]\neq p_{0}(X)'E[q(V)]$. Letting $\Xi(V)\equiv E[p(X)p(X)'|V]$, we show that this is equivalent to $E[q(V)]\in\mathcal{R}(\Xi(V))$
with probability one.

For a matrix $A$, denote by $\mathcal{N}(A)^{\perp}$ the orthogonal
complement of its null space $\mathcal{N}(A)$. The orthogonal complement
of $\mathcal{N}(\Xi(V))$ satisfies $\mathcal{R}(\Xi(V)')=\mathcal{N}(\Xi(V))^{\perp}$
(e.g., \citet[Section 0.6.6]{HJ:2012}), and hence
\[
\mathcal{R}(\Xi(V))=\mathcal{R}(\Xi(V)')=\mathcal{N}(\Xi(V))^{\perp},
\]
by symmetry of $\Xi(V)$. It follows that $E[q(V)]\in\mathcal{R}(\Xi(V))$
with probability one if, and only if, $E[q(V)]\in\mathcal{N}(\Xi(V))^{\perp}$
with probability one. 

By definition of the orthogonal complement, $E[q(V)]\in\mathcal{N}(\Xi(V))^{\perp}$
holds with probability one if, and only if, for each $\Delta(X)\in\mathcal{N}(\Xi(V))$, we have
\begin{equation}
\Delta(X)'E[q(V)]=0\label{eq:lincombikernel}
\end{equation}
with probability one. Adding $p_{0}(X)'E[q(V)]$ on both sides of
(\ref{eq:lincombikernel}), this last condition is equivalent to
\[
[p_{0}(X)+\Delta(X)]'E[q(V)]=p_{0}(X)'E[q(V)],
\]
with probability one for each $\Delta(X)\in\mathcal{N}(\Xi(V))$, and hence there is no observationally equivalent $\overline{p}(X)\neq p_{0}(X)$
such that $\overline{p}(X)'E[q(V)]\neq p_{0}(X)'E[q(V)]$. Therefore, 
$E[q(V)]\in\mathcal{R}(\Xi(V))$ with probability
one if, and only if, the ASF is identified.\qed

\subsection{Proof of Theorem \ref{thm:Theorem6}}

Let $W\equiv1(X\in\mathcal{A})$. Then there exists $\gamma_{J}$
such that for $a_{J}(X)=p_{J}(X)'\gamma_{J}$, we have $E\left[\{a_{J}(X)-W\}^{2}\right]\rightarrow0$,
as $J\rightarrow\infty$. This implies:
\[
E[a_{J}(X)^{2}]\rightarrow E[W^{2}]=E[W]=\Pr(X\in\mathcal{A})>0,
\]
as $J\rightarrow\infty$, and hence, for some small generic constant
$c>0$, we have $E[a_{J}(X)^{2}]\geq c$ for all $J$ large enough.

For all $J$, by the largest eigenvalue condition, 
\[
0<c\leq E[a_{J}(X)^{2}]=\gamma_{J}'E[p_{J}(X)p_{J}(X)']\gamma_{J}\leq\lambda_{\max}(E[p_{J}(X)p_{J}(X)'])||\gamma_{J}||^{2}\leq C||\gamma_{J}||^{2},
\]
and hence $||\gamma_{J}||^{2}\geq c>0$. Therefore, with $\underline{\lambda}(V)\equiv\lambda_{\min}(E[p_{J}(X)p_{J}(X)'|V])$,
\begin{align}
E[a_{J}(X)^{2}|V] & =\gamma_{J}'E[p_{J}(X)p_{J}(X)'|V]\gamma_{J}\nonumber \\
 & \geq\lambda_{\min}(E[p_{J}(X)p_{J}(X)'|V])||\gamma_{J}||^{2}\geq c\underline{\lambda}(V),\label{eq:star}
\end{align}
where $\underline{\lambda}(V)>0$ with probability one, by the conditional
nonsingularity assumption.

Let $\widetilde{\mathcal{V}}$ be the set such that $E[W|V=v]=0$
for all $v\in\widetilde{\mathcal{V}}$. Then, for all $J$,
\begin{align*}
E[\underline{\lambda}(V)1(\widetilde{\mathcal{V}})] & \leq CE[E[a_{J}(X)^{2}|V]1(\widetilde{\mathcal{V}})]\\
 & =CE[|E[a_{J}(X)^{2}|V]-E[W|V]|1(\widetilde{\mathcal{V}})]\\
 & \leq CE[|E[a_{J}(X)^{2}|V]-E[W|V]|]\\
 & \leq CE[|E[a_{J}(X)^{2}|V]-E[W^{2}|V]|]\\
 & =CE[|E[a_{J}(X)^{2}-W^{2}|V]|]\\
 & \leq C|E[E[a_{J}(X)^{2}-W^{2}|V]]|\\
 & =C|E[a_{J}(X)^{2}-W^{2}]|
 \leq CE[\{a_{J}(X)-W\}^{2}]^{\frac{1}{2}}(2\{E[a_{J}(X)^{2}]+E[W^{2}]\})^{\frac{1}{2}},
\end{align*}
where the first inequality is by (\ref{eq:star}), the fourth is by
Jensen's inequality and the last is by Cauchy-Schwarz. Using the facts
that $E[\{a_{J}(X)-W\}^{2}]\rightarrow0$ as $J\rightarrow\infty$,
$E[a_{J}(X)]^{2}<\infty$ and $E[W^{2}]\leq1$, we have established
that $E[\underline{\lambda}(V)1(\widetilde{\mathcal{V}})]=0$ for
$J$ large enough. 

By $\underline{\lambda}(V)>0$ with probability one, $E[\underline{\lambda}(V)1(\widetilde{\mathcal{V}})]=0$
for $J$ large enough implies that $1(\widetilde{\mathcal{V}})=0$.
Therefore, we can conclude that $\Pr(X\in\mathcal{A}|V)=E[W|V]>0$
with probability one, as claimed.\qed

\begin{sloppy}
\subsection{Proof of Corollary \ref{cor:Corollary1}}
By assumption, the marginal support of $X$ is the same as the conditional support of $X$ given $V$, with probability one. Therefore, $\int E[Y|X,V=v]dF_{V}(v)$
is well-defined with probability one, and by $E[Y|X,V]=p(X)'q_0(V)$ under Assumption \ref{ass:Assumption1p}, we have $\int E[Y|X,V=v]dF_{V}(v)=p(X)'E[q_0(V)]$. Thus $p(X)'E[q_0(V)]$ is identified, and hence the ASF also is identified, since $\mu(X)=p(X)'E[q_0(V)]$ in (\ref{eq:model}), under Assumption \ref{ass:DGP} with Assumption \ref{ass:Assumption1p}.
By Theorem \ref{thm:Theorem4}, ASF identification then implies
nonsingularity of $E[p(X)p(X)'|V]$ with probability one. \qed
\par\end{sloppy}

\subsection{Proof of Theorem \ref{thm:Theorem17}}
Let $\widetilde{\mathcal{X}}$ be the set of $X$ values such that $E[q(V)q(V)'|X]$ is nonsingular, and let $\overline{p}(x)\neq p_{0}(x)$
for some value $x$ in $\widetilde{\mathcal{X}}$. With $\lambda(x)\equiv\overline{p}(x)-p_{0}(x) \neq 0$, nonsingularity on  $\widetilde{\mathcal{X}}$ implies
\[
E[\{q(V)'\lambda(X)\}^{2}\mid X=x]=\lambda(x)^{\prime}E[q(V)q(V)'\mid X=x]\lambda(x)>0,
\]
and hence $p_{0}(x)$ is identified from $E[Y|X=x,V]$ on $\widetilde{\mathcal{X}}$.
Therefore, by $E[q(V)]$ known and by the form of the ASF $\mu(X)=p_{0}(X)'E[q(V)]$ under Assumption \ref{ass:DGP} with Assumption \ref{ass:Assumption1q}, 
identification of $p_{0}(X)$ on $\widetilde{\mathcal{X}}$ 
implies identification of the ASF on that set. \qed

\subsection{Proof of Theorem \ref{thm:Theorem8}}

As in the proof  of Theorem 2 in \citet{NeweyStouli:2022}, the matrix
$E[p(X)p(X)'|V]$ is of the form 
\begin{equation}
E[p(X)p(X)'\mid V]=\begin{bmatrix}1 & E[\widetilde{p}(X)'\mid V]\\
E[\widetilde{p}(X)\mid V] & E[\widetilde{p}(X)\widetilde{p}(X)'\mid V]
\end{bmatrix},\label{eq:principal-2}
\end{equation}
and is positive definite if, and only if, the Schur complement of
$1$ in (\ref{eq:principal-2}) is positive definite (\citet[Appendix A.5.5.]{Boyd Vandenberghe 2004}),
i.e., if, and only if, 
\[
E[\widetilde{p}(X)\widetilde{p}(X)'\mid V]-E[\widetilde{p}(X)\mid V]E[\widetilde{p}(X)'\mid V]=\text{var}(\widetilde{p}(X)\mid V),
\]
is positive definite with probability one. By $X=Q_{X|Z}(V|Z)$ with
probability one, we have that $\text{Var}(\widetilde{p}(X)|V)=\text{Var}(\widetilde{p}(Q_{X|Z}(V|Z))|V)$
with probability one. The result then follows by independence of $V$
from $Z$.\qed

\subsection{Proof of Theorem \ref{thm:Theorem9}}

We first state a useful lemma.
\begin{lem}
Suppose $|\mathcal{Z}|<\infty$. $E[p(X)p(X)'|V]$ is nonsingular with probability one only if $\Pr(|\mathcal{Q}(V)|\geq J)=1$.\label{lem:Lemma3} 
\end{lem}

\begin{proof}
By definition of $\mathcal{Z}$ we have that $\Pr(Z=z_{m})\geq\delta>0$
for $m\in\{1,\ldots,|\mathcal{Z}|\}$. Thus, upon using the identity
$X=Q_{X| Z}(V| Z)$ and by independence of $V$ from $Z$, for
$v\in(0,1)$, 
\[
E[p(X)p(X)'\mid V=v]=\sum_{m=1}^{|\mathcal{Z}|}\left\{ p(Q_{X\mid Z}(v\mid z_{m}))p(Q_{X\mid Z}(v\mid z_{m}))^{\prime}\right\} \times\Pr(Z=z_{m}),
\]
is a sum of $|\mathcal{Q}(v)|\leq|\mathcal{Z}|$ rank one $J\times J$
distinct matrices which is singular if $|\mathcal{Q}(v)|<J$. Thus
if $|\mathcal{Q}(V)|<J$ with positive probability, then $E[p(X)p(X)'|V]$
is singular with positive probability. Therefore $E[p(X)p(X)'|V]$
is nonsingular with probability one only if $\Pr(|\mathcal{Q}(V)|\geq J)=1$.
\end{proof}

We now give the proof of Theorem \ref{thm:Theorem9}:

Recall that $\mathcal{Q}(V)$ is the set of distinct values of $z\mapsto Q_{X\mid Z}(V|z)$. 
By assumption $|\mathcal{Z}|<J$, and hence $|\mathcal{Q}(V)|<J$ with probability one. By Lemma \ref{lem:Lemma3} and $J$ finite, this implies that the conditional nonsingularity property does not hold. 
Therefore, by $E[p(X)p(X)']$ nonsingular and Theorem \ref{thm:Theorem4}, the ASF is not identified.   \qed

\subsection{Proof of Theorem \ref{thm:Theorem10}}

We first state a useful lemma.
\begin{lem}
Given a random vector $(Y,W')'$ where $E[Y^{2}]<\infty$ and for
some specified function $m(W)$, we have: $m(W)=E[Y|W]$ if, and only
if, $E[\{Y-m(W)\}a(W)]=0$ for all $a(W)$ with $E[a(W)^{2}]<\infty$.\label{lem:Lemma4}
\end{lem}
\begin{proof}
\begin{sloppy}
If $m(W)=E[Y|W]$ then, for any $a(W)$ with $E[a(W)^{2}]<\infty$,
we have by iterated expectations
\[
E[\{Y-m(W)\}a(W)]=E[\{E[Y|W]-m(W)\}a(W)]=E[0\cdot a(W)]=0.
\]
For the converse result, suppose $E[\{Y-m(W)\}a(W)]=0$ for all $a(W)$
with $E[a(W)^{2}]<\infty$. Choose $a(W)=E[Y|W]-m(W)$. Then, by iterated expectations:
\[
0=E[\{Y-m(W)\}a(W)]=E[\{E[Y|W]-m(W)\}a(W)]=E[\{E[Y|W]-m(W)\}^{2}],
\]
and hence $m(W)=E[Y|W]$.
\end{sloppy}
\end{proof}

We now give the proof of Theorem \ref{thm:Theorem10}:

If $E[\{Y-p(X)'q^{*}(V)\}a(X,V)]=0$ for all $a(X,V)$ with $E[a(X,V)^{2}]<\infty$,
then, by Lemma \ref{lem:Lemma4}, we have $p(X)'q^{*}(V)=E[Y|X,V]$.
For the converse result, suppose $E[Y|X,V]=p(X)'q_{0}(V)$ for some
$q_{0}(V)$. Then, for $\Xi(V)\equiv E[p(X)p(X)'|V]$ and by iterated
expectations, we have:
\begin{align*}
q^{*}(V) & =\Xi(V)^{-1}E[p(X)Y|V]=\Xi(V)^{-1}E[p(X)E[Y|X,V]|V]\\
 & =\Xi(V)^{-1}E[p(X)p(X)'q_{0}(V)|V]=\Xi(V)^{-1}\Xi(V)q_{0}(V)=q_{0}(V).
\end{align*}
Therefore, $E[Y|X,V]=p(X)'q^{*}(V)$, and hence $E[\{Y-p(X)'q^{*}(V)\}a(X,V)]=0$
for all $a(X,V)$ with $E[a(X,V)^{2}]<\infty$, by Lemma \ref{lem:Lemma4}.\qed

\subsection{Proof of Theorem \ref{thm:Theorem11}}

Let $q^{K}(V,b)=(q_{1}^{K}(V,b_{1}),\ldots,q_{J}^{K}(V,b_{J}))'$,
with $q_{j}^{K}(V,b_{j})\equiv\sum_{k=1}^{K}b_{jk}\psi_{k}^{K}(V)$,
$j\in\{1,\ldots,J\}$. 
By Assumption \ref{ass:Assumption1p}, we have $E[Y|X,V]=p(X)'q_{0}(V)$,
where $q_{0}(V)$ is unique with probability one by Theorem \ref{thm:Theorem2}(i).
Thus, for all $b\in\mathbb{R}^{JK}$, 
\begin{align}
E[\{ E[Y|X,V]-b'[p(X)\otimes\psi^{K}(V)]\} ^{2}] & =E[\{ p(X)'q_{0}(V)-b'[p(X)\otimes\psi^{K}(V)]\} ^{2}]\nonumber \\
 & =E[\{ p(X)'[q_{0}(V)-q^{K}(V,b)]\} ^{2}].\label{eq:}
\end{align}
Using that $b_{\textrm{LS}}^{K}$ in (\ref{eq:OLS1}) also satisfies
\[
b_{\textrm{LS}}^{K}=\arg\min_{b\in\mathbb{R}^{JK}}E[\{ E[Y\mid X,V]-b'[p(X)\otimes\psi^{K}(V)]\} ^{2}],
\]
equation (\ref{eq:}) implies that 
\begin{equation}
b_{\textrm{LS}}^{K}=\arg\min_{b\in\mathbb{R}^{JK}}E[\{ p(X)'[q_{0}(V)-q^{K}(V,b)]\} ^{2}].\label{eq:OLS2}
\end{equation}
Thus if, as $K\rightarrow\infty$, 
\[
E[\{ p(X)'[q_{0}(V)-q^{K}(V,b_{\textrm{LS}}^{K})]\} ^{2}]\rightarrow0,
\]
then the result follows.


Define 
\[
\widetilde{b}^{K}\equiv\arg\min_{b\in\mathbb{R}^{JK}}E\left[||q_{0}(V)-q^{K}(V,b)||^{2}\right].
\]
We have that, as $K\rightarrow\infty$, 
\begin{align*}
0\leq E[\{ p(X)'[q_{0}(V)-q^{K}(V,\widetilde{b}^{K})]\} ^{2}] & \leq E[||p(X)||^{2}\,||q_{0}(V)-q^{K}(V,\widetilde{b}^{K})||^{2}]\\
 & =E[E[||p(X)||^{2}|V]\,||q_{0}(V)-q^{K}(V,\widetilde{b}^{K})||^{2}]\\
 & \leq CE[||q_{0}(V)-q^{K}(V,\widetilde{b}^{K})||^{2}]\rightarrow0,
\end{align*}
by Cauchy-Schwarz, iterated expectations, uniform boundedness of $E[||p(X)||^{2}|V=v]$
over $v\in\mathcal{V}$ and Assumption \ref{ass:Assumption3-sieve}.
Thus $\widetilde{b}^{K}$ is a minimizer of (\ref{eq:OLS2}) for $K$
large enough. We have that $E[p(X)p(X)'|V]$ is nonsingular with probability
one by Assumption \refp{ass:cond_nonsingularity}, and that $E[\psi^{K}(V)\psi^{K}(V)']$ is nonsingular by Assumption \ref{ass:Assumption3-sieve}. 
Thus the matrix $E[\{p(X)\otimes\psi^{K}(V)\}\{p(X)\otimes\psi^{K}(V)\}']$
is nonsingular for each $K$ by Theorem 2 in \citet{NeweyStouli2021}.
Therefore, $\widetilde{b}^{K}$ is the unique minimizer of (\ref{eq:OLS2})
for $K$ large enough. Conclude that $b_{\textrm{LS}}^{K}=\widetilde{b}^{K}$
for $K$ large enough, and the result follows. \qed

\end{document}